\documentclass[twocolumn,twocolappendix]{aastex631}

\usepackage{graphicx}	
\usepackage{amsmath}	
\usepackage{amssymb}	
\usepackage{multirow}
\usepackage{xcolor}
\definecolor{black}{rgb}{0,0,0}
\definecolor{red}{rgb}{1,0,0}
\definecolor{green}{rgb}{0,0.5,0}
\newcommand\bla{\color{black}}

\shorttitle{IRAS 18314--0720}
\shortauthors{Anindya Saha et al.}

\begin{document}
\title{Direct observational evidence of multi-epoch massive star formation in G24.47+0.49}
\correspondingauthor{Anindya Saha}
\email{anindya.s1130@gmail.com}
\correspondingauthor{Anandmayee Tej}
\email{tej@iist.ac.in}
\correspondingauthor{Hong-Li Liu}
\email{hongliliu2012@gmail.com}
\author{Anindya Saha}
\affiliation{Indian Institute of Space Science and Technology, Thiruvananthapuram 695 547, Kerala, India}
\author{Anandmayee Tej}
\affiliation{Indian Institute of Space Science and Technology, Thiruvananthapuram 695 547, Kerala, India}
\author{Hong-Li Liu}
\affiliation{Department of Astronomy, Yunnan University, Kunming, 650091, People's Republic of China}
\author{Tie Liu}
\affiliation{Shanghai Astronomical Observatory, Chinese Academy of Sciences, 80 Nandan Road, Shanghai 200030, People's Republic of China}
\affiliation{Key Laboratory for Research in Galaxies and Cosmology, Shanghai Astronomical Observatory, Chinese Academy of Sciences, 80 Nandan Road, Shanghai 200030, People's Republic of China}
\author{Guido Garay}
\affiliation{Departamento de Astronom\'{\i}a, Universidad de Chile, Las Condes, Santiago 7550000, Chile}
\author{Paul F. Goldsmith}
\affiliation{Jet Propulsion Laboratory, California Institute of Technology, 4800 Oak Grove Drive, Pasadena, CA 91109, USA}
\author{Chang Won Lee}
\affiliation{University of Science and Technology, Korea (UST), 217 Gajeong-ro, Yuseong-gu, Daejeon 34113, Republic of Korea}
\affiliation{Korea Astronomy and Space Science Institute, 776 Daedeokdae-ro, Yuseong-gu, Daejeon 34055, Republic of Korea}
\author{Jinhua He}
\affiliation{Yunnan Observatories, Chinese Academy of Sciences, Phoenix Mountain, East Suburb of Kunming, 650216, Yunnan, People's Republic of China}
\affiliation{Chinese Academy of Sciences, South America Center for Astrophysics (CASSACA) at Cerro Cal\'{a}n, Camino El Observatorio \#1515, Las Condes, Santiago, Chile}
\affiliation{Departamento de Astronom\'{\i}a, Universidad de Chile, Las Condes, Santiago 7550000, Chile}
\author{Mika Juvela}
\affiliation{Department of Physics, P.O. box 64, FI- 00014, University of Helsinki, Finland}
\author{Leonardo Bronfman}
\affiliation{Departamento de Astronom\'{\i}a, Universidad de Chile, Las Condes, Santiago 7550000, Chile}
\author{Tapas Baug}
\affiliation{Satyendra Nath Bose National Centre for Basic Sciences, Block-JD, Sector-III, Salt Lake, Kolkata-700 106, India}
%
%
\author{Enrique Vázquez-Semadeni}
\affiliation{Instituto de Radioastronomía y Astrofísica, Universidad Nacional Autónoma de México, Antigua Carretera a Pátzcuaro 8701, Ex-Hda. San José de la Huerta, 58089 Morelia, Michoacán, México}
\author{Patricio Sanhueza}
\affiliation{National Astronomical Observatory of Japan, National Institutes of Natural Sciences, 2-21-1 Osawa, Mitaka, Tokyo 181-8588, Japan}
\affiliation{Astronomical Science Program, The Graduate University for Advanced Studies, SOKENDAI, 2-21-1 Osawa, Mitaka, Tokyo 181-8588, Japan}
\author{Shanghuo Li}
\affiliation{Max Planck Institute for Astronomy, Königstuhl 17, D-69117 Heidelberg, Germany}
\author{James O. Chibueze} %
\affiliation{Department of Mathematical Sciences, University of South Africa, Cnr Christian de Wet Rd and Pioneer Avenue, Florida Park, 1709, Roodepoort, South Africa}
\affiliation{Department of Physics and Astronomy, Faculty of Physical Sciences, University of Nigeria, Carver Building, 1 University Road, Nsukka 410001, Nigeria}
\author{N. K. Bhadari}
\affiliation{Physical Research Laboratory, Navrangpura, Ahmedabad 380009, India}
\author{Lokesh K. Dewangan} %
\affiliation{Physical Research Laboratory, Navrangpura, Ahmedabad 380009, India}
\author{Swagat Ranjan Das} %
\affiliation{Departamento de Astronom\'{\i}a, Universidad de Chile, Las Condes, Santiago 7550000, Chile}
\author{Feng-Wei Xu}
\affiliation{Kavli Institute for Astronomy and Astrophysics, Peking University, Beijing 100871, People's Republic of China}
\affiliation{Department of Astronomy, School of Physics, Peking University, Beijing, 100871, People's Republic of China}
\affiliation{I. Physikalisches Institut, Universität zu Köln, Zülpicher Str. 77, D-50937 Köln, Germany}
\author{Namitha Issac}
\affiliation{Shanghai Astronomical Observatory, Chinese Academy of Sciences, 80 Nandan Road, Shanghai 200030, People's Republic of China}
\author{Jihye Hwang}
\affiliation{Korea Astronomy and Space Science Institute, 776 Daedeokdae-ro, Yuseong-gu, Daejeon 34055, Republic of Korea}
\author{L. Viktor Tóth}
\affiliation{Institute of Physics and Astronomy, Eötvös Lorànd University, Pázmány Péter sétány 1/A, H-1117 Budapest, Hungary}
\affiliation{University of Debrecen, Institute of Physics, H-4026, Debrecen, Bem ter 18}
%
\begin{abstract}
Using new continuum and molecular line data from the ALMA Three-millimeter Observations of Massive Star-forming Regions (ATOMS) survey and archival VLA, 4.86~GHz data, we present direct observational evidence of hierarchical triggering relating three epochs of massive star formation in a ring-like H\,{\small II} region, G24.47+0.49. We find from radio flux analysis that it is excited by a massive star(s) of spectral type O8.5V–O8V from the first epoch of star formation. 
The swept-up ionized ring structure shows evidence of secondary collapse, and within this ring a burst of massive star formation is observed in different evolutionary phases, which constitutes the second epoch. 
ATOMS spectral line (e.g., HCO$^+$(1--0)) observations reveal an outer concentric molecular gas ring expanding at a velocity of $\sim$ 9$\,\rm km\,s^{-1}$, constituting the direct and unambiguous detection of an expanding molecular ring. 
It harbors twelve dense molecular cores with surface mass density greater than 0.05$\,\rm g\,cm^{-2}$, a threshold typical of massive star formation. Half of them are found to be subvirial, and thus in gravitational collapse, making them third  epoch of potential massive star-forming sites.

\end{abstract}
%
\section{Introduction} 
\label{sec:Introduction}
Massive stars ($M_{\star} \gtrsim 8\,\rm M_{\odot}$), with their powerful mechanical and radiative feedback, play a crucial role in regulating star formation within their natal environments. They can either initiate the formation of a subsequent 
generation of stars \citep[e.g.,][]{Churchwell2006} or disperse the surrounding molecular gas, consequently inhibiting further star formation \citep[e.g.,][]{Walch2012,Pabst2019,Bonne2023}. H\,{\small II} regions and their role in triggered star formation has been in
focus, since the pioneering work by \citet{ElmegreenLada1977}.
Over the last decade or so, there has been a plethora of observational evidences linking the expansion of H\,{\small II} regions to triggered star formation \citep[e.g.,][and references therein]{Zavagno2006,Zavagno2007,Figueira2017}. The peripheries of infrared dust bubbles \citep{Churchwell2006,Kendrew2012} have served as ideal sites to investigate triggered star formation through various competing mechanisms \citep[e.g.,][]{Deharveng2010,Thompson2012,Kendrew2012,Hongliliu2016,Swagat2017,Bhadari2021,Zhang2023ATOMSXIII}. However, theoretically \citep[e.g.][]{Dale2015,Semadeni2020} and observationally \citep[e.g.][]{Cambresy2013}, it is also seen that commonly used signposts of triggered star formation do not always lead to
definite conclusions on positive feedback. Hence, these need to be cautiously interpreted to better constrain the impact of stellar feedback on triggering star formation.%

\begin{figure*}
    \centering
    \includegraphics[width=0.9\textwidth]{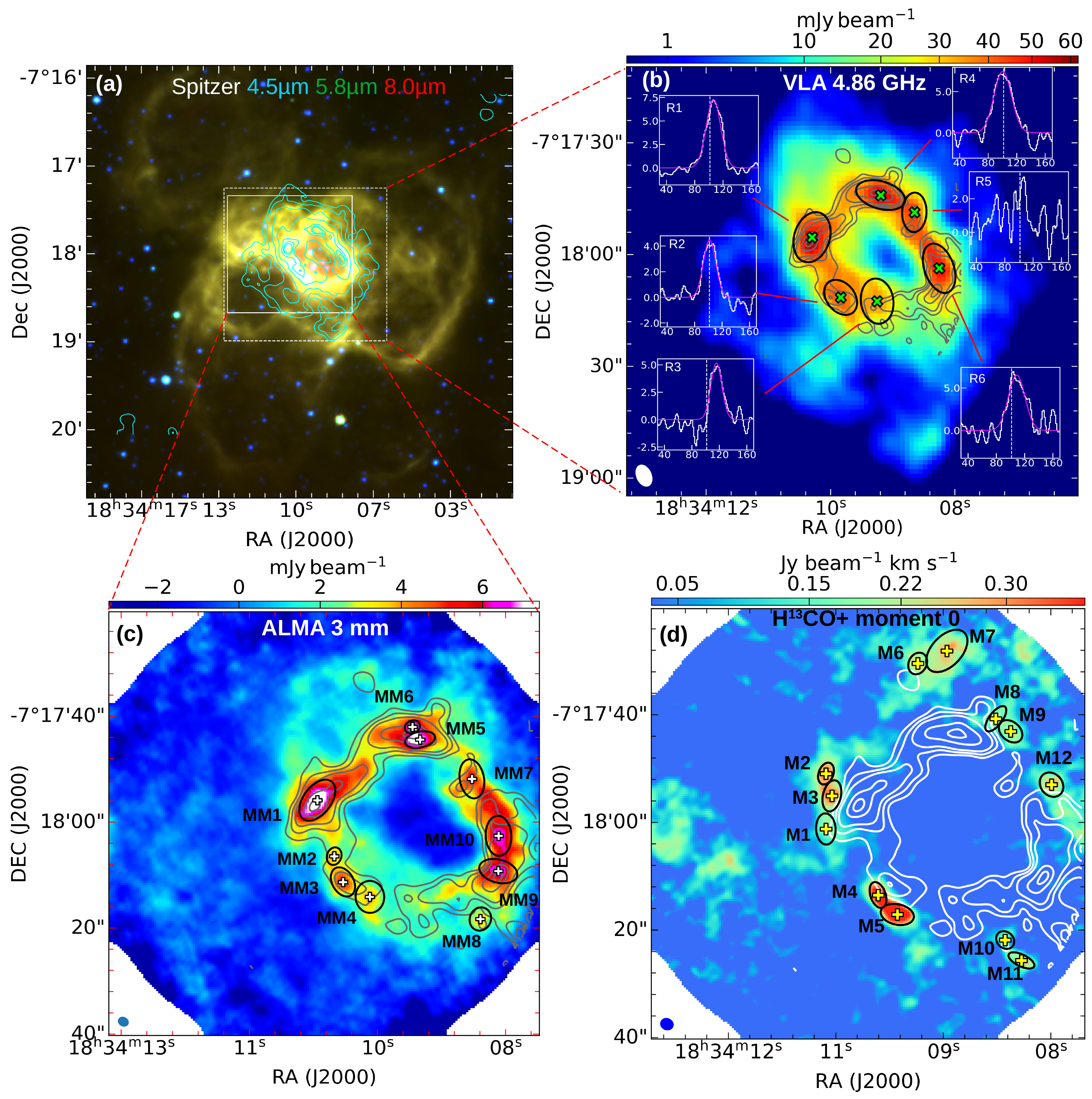}
    \caption{ Morphology of the region associated with G24.47. (a) {\it Spitzer}-IRAC colour composite image overlaid with VLA 4.86 GHz contours at 3, 7, 20, 50, 90, 110, 120 times $\sigma$ (= 0.4 $\rm mJy\,beam^{-1}$). (b) VLA 4.86~GHz map overlaid with ATOMS H40$\alpha$ contours (Gaussian smoothed over 5 pixels) starting at 2$\sigma$ (= 0.04 $\rm Jy\,beam^{-1}\,km\,s^{-1}$) in steps of 1$\sigma$. The displayed ellipses are identified VLA cores with their central positions (X) marked. The insets show the H40$\alpha$ spectra (boxcar smoothed by four channels; velocity resolution of 6.0 $\rm km\,s^{-1}$) of the cores along with their respective Gaussian fits.  The LSR velocity of 101.5 $\rm km\,s^{-1}$ is indicated in each. (c) ALMA 3~mm map overlaid with the H40$\alpha$ contours displayed in (b). The displayed ellipses are identified 3~mm cores with their central positions (+) marked. (d) Moment zero map (3-pixel smoothed using Gaussian kernel) of H$^{13}$CO$^+$ in the velocity range 93.0 to 113.0 $\rm km\,s^{-1}$. The displayed ellipses are identified molecular cores with their central positions (+) marked. Ellipses drawn in the lower left corner of (b), (c), and (d) represent the beams of the respective maps displayed.}
    \label{fig:RGB_vla_ha}
\end{figure*}
%

These observational studies have offered a comprehensive insight into triggered star formation, connecting two generations of stars. In comparison, observational evidence for hierarchical triggering and multi-generation star formation is still scarce \citep[e.g.,][]{Oey2005,Purcell2009,Areal2020}.
\citet{Oey2005} suggest a three-generation system of hierarchically triggered star formation in the W3/W4 complex, where expanding superbubbles and mechanical feedback from massive stars initiate later generations of star formation. 
Similarly, \citet{Purcell2009} report multi-generations of massive star formation in the NGC 3576, which is embedded in the center of an extended filamentary cloud. Here the expansion of the H\,{\small II} region into the ambient molecular cloud leads to the formation of high-mass stars along the dusty filament. 
In a recent work, based on the spatial and temporal correspondences derived in their analysis, \citet{Areal2020} propose three generations of star formation associated with a massive star LS II +26 8.

In this paper, we investigate the H\,{\small II} region, G24.47+0.49 (hereafter G24.47), likely ionized by an early O-type star \citep{Garay1993}, to probe possible signatures of hierarchical triggering and multi-epoch star formation. Observed as part of several radio surveys \citep{Wink1982,Lockman1989,Churchwell1990,Garay1993,Becker1994,Walsh1998}, G24.47 is associated with IRAS 18314--0720 and is located at a distance of 5.82~kpc \citep{Urquhart2018}. This source is also associated with the massive (6095$\,\rm M_{\odot}$) ATLASGAL clump, AGAL024.471+00.487 \citep{Urquhart2018}. The 4.5, 5.8, and 8.0 $\mu$m colour-composite {\it Spitzer}-IRAC\footnote{Images taken from the archives of the Galactic Legacy Infrared Midplane Survey Extraordinaire \citep[{GLIMPSE;}][]{Benjamin2003}.} image illustrated in Figure \ref{fig:RGB_vla_ha}(a) presents an interesting morphology, where G24.27 reveals as a bright ring located at the center of a complex region displaying bubble-like structures, pillars, and arcs.

The paper is organized as follows. Section \ref{sec:obs_data} discusses the ALMA observations carried out as part of the ALMA Three-millimeter Observations of Massive Star-forming Regions (ATOMS) survey and the other multi-wavelength archival data used in this study. Results obtained from the dust continuum, ionized emission, and molecular line analysis are presented in Section \ref{sec:3mm-vla-mol}. Discussion on the three observed epochs of star formation is presented in Section \ref{sec:discussion}, and the overall picture of hierarchical triggering and multi-epoch star formation in G24.47 is discussed in \ref{sec:Hierarchical-trigger-multi-epoch}. Section \ref{sec:conclusion} summarizes the results. 
%
\section{Observations and archival data}
\label{sec:obs_data}
For this study, we have utilized data from the ATOMS survey and other archival datasets. Brief descriptions of these are given in the following subsections.
\subsection{ALMA observations}
G24.47 was observed as part of the ATOMS survey (Project ID: 2019.1.00685.S; PI: Tie Liu), which aims to study 146 massive star-forming clumps. Details of the survey can be found in \citet{Liu2020ATOMSI}. The 12-m + 7-m combined data for continuum and line emission are used here. The maps have a field of view of ~80$\arcsec$ or 2.26 pc at the distance of G24.47, and a maximum recoverable scale of 76$\arcsec$.2 or 2.15 pc. To probe the kinematics and dynamics of the associated ionized and dense gas, we use the H40$\alpha$ hydrogen radio recombination line (RRL) along with H$^{13}$CO$^+$ (1-0) and HCO$^+$ (1-0) molecular line transitions. The synthesized beam size for the continuum and H40$\alpha$ is 2$\arcsec$.1 $\times$ 1$\arcsec$.8. For molecular line observations, they are 2$\arcsec$.4 $\times$ 2$\arcsec$.1 and 2$\arcsec$.3 $\times$ 2$\arcsec$.0 for H$^{13}$CO$^+$ and HCO$^+$, respectively. The \textit{rms} noise is $\sim$0.65 mJy $\rm beam^{-1}$ for the continuum, and $\sim$[3.5, 8.5, 12] mJy $\rm beam^{-1}$ for the [H40$\alpha$, H$^{13}$CO$^+$, HCO$^+$] lines at the native velocity resolution of [1.5, 0.2, 0.1] $\rm km\,s^{-1}$. 
\subsection{Archival data}
To probe the radio continuum emission associated with this region, we use VLA archival data at 4.86~GHz. The observations were conducted on 3 March 1988 using the VLA C configuration\footnote{\url{https://science.nrao.edu/facilities/vla/docs/manuals/propvla/array_configs}} (Legacy ID: AB414; R. Becker). The image is retrieved from National Radio Astronomy Observatory VLA Archive Survey\footnote{The NVAS can be browsed through  \url{http://www.vla.nrao.edu/astro/nvas/}} (NVAS) which has a beam size of 5$\arcsec$.9 × 3$\arcsec$.9 and a \textit{rms} noise of 0.4 $\rm mJy\,beam^{-1}$.
To identify the ionizing source associated with the H {\small II} region, G24.47, we use the near-infrared (NIR) \textit{JHK} photometric data for point sources from the Two Micron All Sky Survey \citep[2MASS;][]{Skrutskie2006} and UKIRT Infrared Deep Sky Survey \citep[UKIDSS;][]{Lawrence2007}, which were taken during the UKIDSS Galactic Plane Survey \citep[GPS;][data release 6]{Lucas2008}.
The angular resolution of 2MASS and UKIDSS data are $\sim$2\arcsec and 0.9\arcsec, respectively.
%
\section{Results}
\label{sec:3mm-vla-mol}
\subsection{4.86~GHz and 3~mm continuum emission}
VLA 4.86~GHz and ATOMS 3~mm continuum maps are shown in Figure \ref{fig:RGB_vla_ha}. 
The VLA map displays a distinct ring morphology (radius $\sim$0.8~pc) 
of bright, ionized gas emission with prominent peaks, a low-emission inner region displaying an almost empty cavity like structure at the center, and extended faint emission beyond the ring. Such ring-like morphology of H {\small II} regions could be  associated with flat, sheet-like parental cloud structures \citep{Beaumont2010,Kabanovic2022}. 
The inner rim of the ring is observed to be dominated by 8.0~$\mu$m emission (see Figure \ref{fig:RGB_vla_ha}(a)). 
This could be attributed to thermal dust emission from the forming hot massive stars in the ring or emission from polycyclic aromatic hydrocarbons \citep[e.g.,][]{Watson2008}, which are indicative of photodissociation regions.  
The ATOMS 3~mm continuum and H40$\alpha$ line emission are also seen to closely trace the bright ring structure (see Figure \ref{fig:RGB_vla_ha}(b) and (c)).

Both the 4.86 GHz and 3~mm maps reveal the presence of compact cores in the bright ring. To understand the nature of these cores, we implement the approach followed by \citet{Saha2022} and use a combination of the \texttt{DENDROGRAM} algorithm and the \texttt{CASA} imfit task to extract the cores. In total, six radio (R1 - R6) and ten 3~mm (MM1 - MM10) cores are identified. The retrieved core apertures and the corresponding peak positions are shown in Figures \ref{fig:RGB_vla_ha}(b) and (c). The details of the procedure followed and parameters used are elaborated in Appendix \ref{apndx:core_extract}. 
 
To characterize the radio cores, we calculate the physical parameters such as the emission measure ($EM$), number of Lyman continuum photons emitted per second ($N_{\rm Ly}$), and electron density ($n_{\rm e}$) using Equations 6-8 from \citet{Schmiedeke2016}. For this, we assume the 4.86~GHz emission to be optically thin, and adopt the value of the electron temperature to be 6370~K from \citet{Quireza2006}. The estimated parameters are tabulated in Table \ref{tab:param-VLAcores}. 

Of the ten 3~mm cores identified in our study, only the brighter ones (MM1, MM5, and MM6) were detected by \citet{Liu2021ATOMSIII} using higher resolution 12-m array ATOMS data. Similar issue is also discussed in \citet{Sanhueza2019}, where the inclusion of more extended emission from a more compact configuration results in 20\% increase in the number of cores detected.
None of these cores show any emission in the molecular line transitions of H$^{13}$CO$^+$ and HCO$^+$. This can be inferred from Figures \ref{fig:RGB_vla_ha}(c) -(d), where the 3~mm ring is seen to be mostly devoid of H$^{13}$CO$^+$ emission and also similar distribution is seen for HCO$^+$ emission (see Figure \ref{fig:moment-maps} (a)). 
This suggests that the 3~mm continuum emission is predominantly free-free emission \citep{Keto2008,Zhang2023ATOMSXIV} with appreciably less contribution from cold dust emission. We verify this by following the method described in \citet{Liu2023} and find six of the 3~mm cores to have more than 50\% contribution from free-free emission. However, the existence of hot dust associated with these cores is evident from the presence of MIR emission shown in Figure \ref{fig:RGB_vla_ha}(a). Table \ref{tab:parameter3mm} lists the estimated core parameters.  

\subsection{Molecular line emission}
Figure \ref{fig:moment-maps}(a) shows the distribution of HCO$^+$ molecular line emission. 
The molecular line emission encircles the H40$\alpha$ and 3~mm ring. Henceforth, we refer to this as the molecular gas ring, the morphology of which is similar to the ionized gas ring as traced by radio 4.86\,GHz emission.
The molecular ring also shows the presence of bright, compact cores.
Considering the H$^{13}$CO$^+$ (1-0) line to be optically thin \citep[e.g.][]{sanhueza2012,Saha2022}, we utilize the velocity integrated intensity
(i.e., moment zero) map of this transition to extract the dense molecular cores.
The same procedure as used for extraction of the radio and 3~mm cores is implemented (see Appendix \ref{apndx:core_extract} for details) and ten molecular cores are identified. A careful visual inspection shows the presence of two additional cores that were not detected from this map. For these, we retrieve the parameters using the same approach on the column density map (see Appendix \ref{apndx:column_density}).
The cores are labelled, M1 - M12 of which M1 and M10 are extracted from the column density map.
The retrieved apertures are drawn in Figures \ref{fig:RGB_vla_ha}(d) and \ref{fig:cdmap_molcore}.
Core masses are calculated from the generated hydrogen column density map using,
\begin{equation}
    M = \mu_{\rm H_2} m_{\rm H} A \sum N({\rm H_{2}}), 
\end{equation}
where $\mu_{\rm H_2}$, $ m_{\rm H}$ are the mean molecular weight and mass of the hydrogen atom, respectively, $A$ is the pixel area, and $\sum N(\rm H_{2})$ is the sum of the column density values for the pixels in the core area.

Next, to examine the gravitational stability of the cores, we estimate the virial parameter ($\alpha_{\rm vir}$), which represents the ratio of the virial mass ($M_{\rm vir}$) to the mass of the individual cores ($M_{\rm core}^{\rm mol}$). $M_{\rm vir}$ is given by \citep{Contreras2016}
\begin{align}
    M_{\rm vir} &= \frac{5\ R_{\rm eff}^{\rm mol}\ \Delta V^2}{8\ {\rm ln}(2)\ a_1\ a_2\ G} \nonumber\\
    &\sim 209\ \frac{1}{a_1\ a_2} \left(\frac{\Delta V}{\rm km\ s^{-1}} \right)^2\ \left(\frac{r}{\rm pc}\right) \rm M_{\odot},
\label{eqn:M_vir}
\end{align}
where $R_{\rm eff}^{\rm mol}$ is the effective radius of the core and $\Delta V$ is the line width of the fitted Gaussian profiles to the observed H$^{13}$CO$^+$ spectra. In the presence of two velocity components, we calculated the average of the line widths obtained from fitting each component individually, following the approach used in \citet{Saha2022}.
The constant $a_1$ accounts for the correction for power-law density distribution. It is given as $ a_1 = (1-p/3)/(1-2p/5)$ for $ p < 2.5$ \citep{Bertoldi1992}, where we adopt $p=1.8$ \citep{Contreras2016}. The constant $a_2 = (arcsin\,e)/e$ takes into account the shape of the core, $e$ being the eccentricity. The estimated parameters of the molecular cores are listed in Table \ref{tab:molcore-param}.
\subsection{Velocity structure of G24.47}
\label{sec:vel-struc-ring}
%
\begin{figure*}
    \centering
    \includegraphics[width=\linewidth]{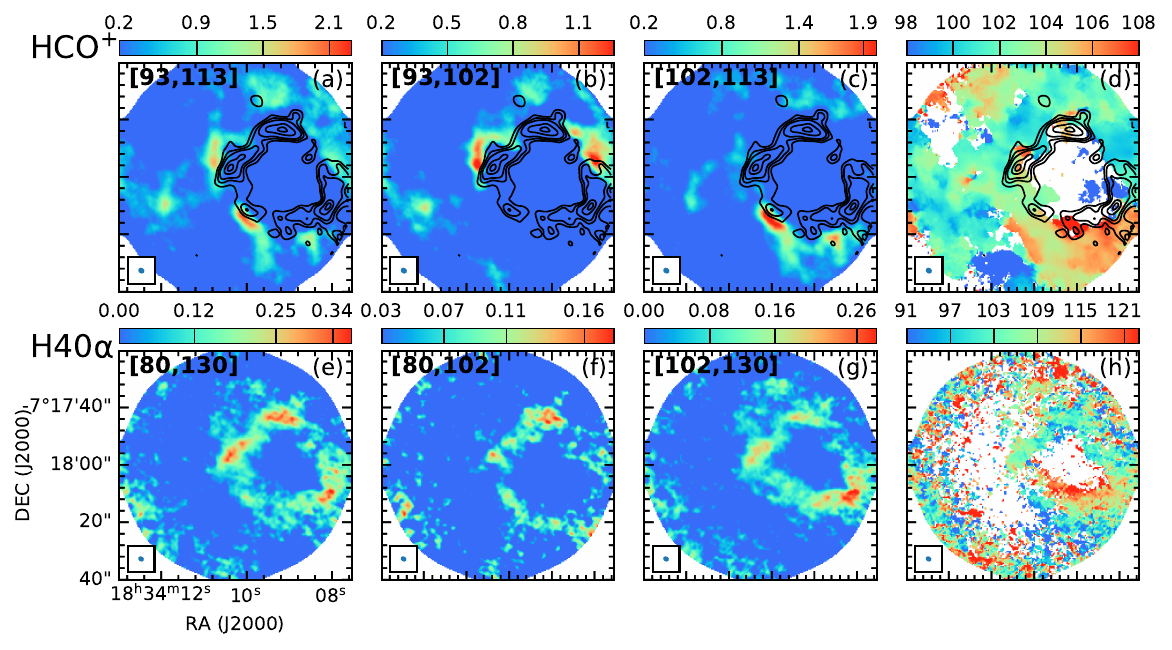}
    \caption{Moment zero (left three) and moment one (rightmost) maps of HCO$^+$ and H40$\alpha$ observed towards G24.47 are shown in panels (a) -- (d) and (e) -- (h), respectively. The velocity ranges used to obtain the moment zero maps are given in the top left of each panel. The colour bar indicates the flux scale in $\rm Jy\,beam^{-1}\,km\,s^{-1}$ and $\rm km\,s^{-1}$ for moment zero and moment one maps, respectively. 
    The overlaid contours (in panels (a) -- (d)) show the H40$\alpha$ emission (presented as colorscale in panel (e)) with contour levels starting at 2$\sigma$ ( $\sigma =$ 0.04 $\rm Jy\,beam^{-1}\,km\,s^{-1}$) in steps of 1$\sigma$. 
    These contours are smoothed over five pixels using Gaussian kernel. The moment zero maps (in panels (e) -- (g)) are smoothed across three pixels using Gaussian kernel.
    The beam is indicated at the bottom left corner in each panel.}
    \label{fig:moment-maps}
\end{figure*}
\begin{figure}
    \centering
    \includegraphics[width=0.94\linewidth]{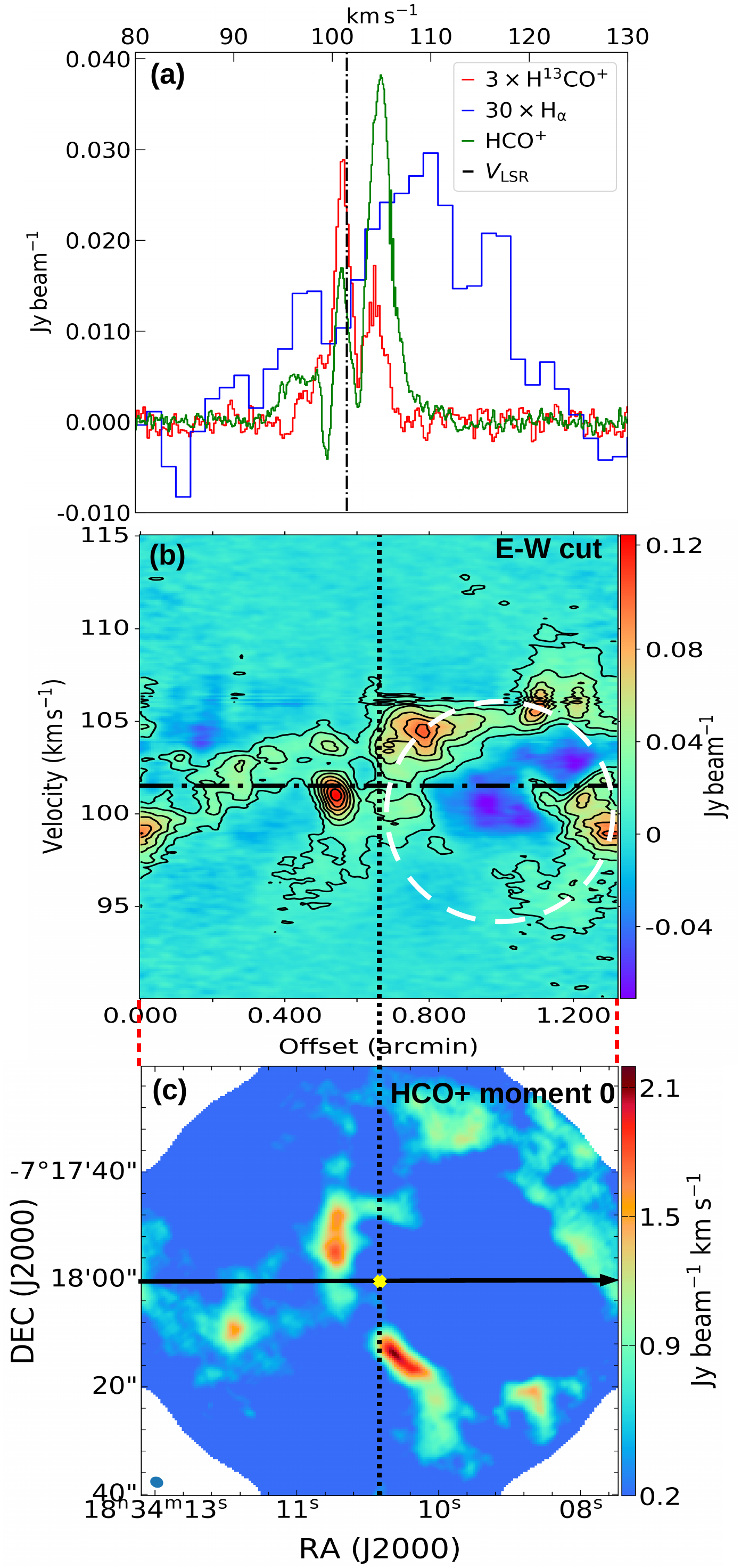}
    \caption{(a) Spectra (averaged over entire molecular ring) of H$^{13}$CO$^+$ (1-0), HCO$^{+}$ (1-0) and H40$\alpha$ towards G24.47 are shown in red, green and blue lines, respectively. 
    H$^{13}$CO$^+$ (1-0) and H40$\alpha$ spectra are scaled-up by factors of 3 and 30, respectively. H40$\alpha$ spectrum is boxcar smoothed by four channels, resulting in a velocity resolution of 6.0 $\rm km\,s^{-1}$. 
    (b) PV diagram of HCO$^+$ along the PV cut in east-west direction, centered on the ALMA phase center (marked in panel (c)). The contours start at 3$\sigma$ ($\sigma = 3.0\,\rm mJy\,beam^{-1}$) in steps of 6$\sigma$. The circular velocity structure is indicated by the white dashed line. 
    The black dash-dot line marks the LSR velocity of 101.5 $\rm km\,s^{-1}$ in panels (a) and (b).
    (c) Moment zero map of HCO$^+$, this panel is same as Figure \ref{fig:moment-maps}(a).}
    \label{fig:pv-diag}
\end{figure}

To inspect and compare the spatial distribution of molecular and ionized gas emission, we created separate moment zero maps of HCO$^{+}$ and H40$\alpha$ over three velocity extents, (i) full velocity range (ii) blue-shifted velocity range ($<102\,\rm km\,s^{-1}$ to 80 or 93$\,\rm km\,s^{-1}$) and (iii) red-shifted velocity range ($>102\,\rm km\,s^{-1}$ up to 113 or 130$\,\rm km\,s^{-1}$). The moment zero maps are shown in Figures \ref{fig:moment-maps}(a) - (c) and (e) - (g). 
The above velocity ranges are estimated from the average spectra plotted of these line transitions presented in  Figure \ref{fig:pv-diag}(a). The H$^{13}$CO$^+$ and HCO$^+$ line profiles show prominent peaks around 101.0 and 104.5$\,\rm km\,s^{-1}$. On the other hand, 
the RRL shows a broader profile with a distinct peak at approximately 110.0$\,\rm km\,s^{-1}$. The systemic velocity of G24.47 is 101.5$\,\rm km\,s^{-1}$ \citep{Schlingman2011, Urquhart2018}.

As seen from the HCO$^+$ moment zero maps, in the velocity range [93, 102]$\,\rm km\,s^{-1}$, the northern part of the molecular gas ring is visible, whereas in the 102 to 113$\,\rm km\,s^{-1}$ range, the emission traces the southern part of the ring. Channel maps (in 1$\,\rm km\,s^{-1}$ bins) are also presented in Appendix A, illustrating the above morphology. Similar morphology is also evident from the H$^{13}$CO$^+$ transition, where the diffuse emission is less pronounced.
Figure \ref{fig:moment-maps}(d) presents the intensity-weighted mean velocity (i.e., moment one) map that conforms with the above velocity structure. The northern portion of the molecular ring is blue-shifted and the southern part red-shifted, suggesting expansion of the molecular gas ring. 
In comparison, the complete ring morphology can be discerned over the entire velocity range for the H40$\alpha$ emission, though the moment one map (Figure \ref{fig:moment-maps}(h)) reveals a distinct velocity gradient, similar to that seen in HCO$^{+}$.

To examine the velocity field in more detail, we construct position-velocity (PV) diagram of HCO$^{+}$ towards G24.47 along the east-west direction (offset increases from east to west direction). The PV diagram is shown in Figure \ref{fig:pv-diag}(b).
The integrated intensity map is also included (Figure \ref{fig:pv-diag}(c)) for easy correlation with the location of the velocity components. Consistent with the moment one map, the PV diagram also displays signatures of expansion. Velocity differences are evident along the PV cut. Additionally, it displays an almost circular velocity pattern, that is in very good agreement with the results of simulations of expanding shells discussed in the literature \citep[e.g.,][]{Arce2011,Wang2016}. 
The PV diagram along the north-south direction (not presented here), also shows evidence of expansion but not as prominent.
Following the approach outlined in \citet{Arce2011}, we obtain a rough estimate of the expansion velocity to be $\rm \sim9\,km\,s^{-1}$ by considering the maximum red- and blue-shifted velocities observed. This is consistent with the velocity shifts seen in the moment one map (see Figure \ref{fig:moment-maps}(d)). Given the lower velocity resolution and signal-to-noise ratio of the H40$\alpha$ emission, it was not possible to evaluate the expansion, if any, of the ionized ring.

\section{Discussion}
\label{sec:discussion}
In this section, we probe hierarchical triggering and multi-epoch massive star formation scenario in G24.47.
\subsection{H\,{\small II} region G24.47}
In unveiling the multi-epoch star formation in G24.47, the first one is the massive star(s) responsible for the creation of the H\,{\small II} region. Based on their 1.5~GHz radio flux density, \citet{Garay1993} proposed a massive ionizing star of spectral type O5.5. We note here that these authors have used the far distance in their analysis. We re-visit this estimation using the 4.86~GHz VLA map. 
Integrating within the $3\sigma$ contour, and subtracting the contribution from the six detected compact radio cores, the flux density is calculated to be $\sim$0.9~Jy which translates to a Lyman-continuum photon flux, $N_{\rm Ly}$, of $\sim$3.4$\times\,10^{48}\,\rm s^{-1}$. Assuming a single star to be responsible for the ionization of G24.47, and comparing the calculated $N_{\rm Ly}$ with that of early type main-sequence stars tabulated in \citet{Panagia1973}, we infer the spectral type to be O8.5V - O8V. However, this can only be considered as a lower limit, since dust absorption of Lyman continuum photons is not accounted for here, which can be significant as shown by many studies \citep[e.g.,][]{Paron2011}. 
The central cavity of G24.47, which is observed to be mostly dust-free and devoid of molecular line emission, is likely to be carved out by the powerful radiation and wind of the massive star(s).  
We attempt to identify the ionizing star(s) by studying the stellar population located within the observed ionized gas emission. This is carried out using NIR colour-magnitude and colour-colour plots \citep[e.g.,][]{Potdar2022}. For our study, we have used 2MASS and UKIDSS datasets. The detailed procedure is discussed in Appendix \ref{apndx:Ionizing_star}. Following this, 12 candidate massive (with spectral type earlier than B3), Class III stars, namely, E1 - E12, are identified within the VLA radio emission. The location of these are shown in Figure \ref{fig:cc-cm}(c). The positions and $JHK$ magnitudes of these sources are listed in Table \ref{tab:ionisingstar-param}.

By a simple argument, the symmetry of the ionized ring morphology suggests a centrally located ionizing star or a group of ionizing stars. The sources, E10, E11, and E12 are located at the center of the cavity. However, their spectral types from the colour-magnitude plot lie between $\sim$ B3$-$B0.5. So the individual or the sum of their Lyman-continuum photon flux is not consistent with that calculated from the observed 4.86~GHz flux density. The other identified early type stars are mostly located on the bright, ionized ring. These could be bonafide ionizing sources, since it is possible that the ionizing star is displaced from the center due to its proper motion. Such a geometry is observed in the Orion Veil bubble, where the exciting massive star, $\rm \theta^1\,Ori\,C$ is seen offset from the center \citep[see Fig. 3 of][]{Pabst2019}.    
Furthermore, simulations of expanding H\,{\small II} regions, discussed in \citet{Maclow2007} and \citet{Hunter2008}, also show the possibility of formation of nearly spherical shells with off-center ionizing source. Under this scenario, sources E2, E3, E4, and E8 are potential candidates. The NIR spectral type inferred from the colour-magnitude diagram is reasonably consistent with the estimated radio spectral type. The above analysis, however, restricts any conclusive identification of the ionizing star of G24.47.  
 
\subsection{The inner ring: radio and 3~mm continuum emission}
Six radio cores are identified in this inner ring of ionized gas. 
Summarizing the results, we find the radius ($R_{\rm core}^{\rm VLA}$), $EM$, $n_{\rm e}$, and $M_{\rm ion}$ in the range of $\sim$ [0.1, 0.2] pc, [1.6, 2.5]$\times 10^6\,\rm cm^{-6}\,pc$, [2.3, 3.0]$\times 10^3\,\rm cm^{-3}$, and [0.5, 1.2]$\,\rm M_\odot$, respectively.
In the same order, the median values are estimated to be 0.28 pc, $2.2\times 10^6\,\rm cm^{-6}\,pc$, 2.8$\times 10^3\,\rm cm^{-3}$, and 0.8$\,\rm M_\odot$.
The derived radio properties of these compact VLA cores are consistent with those of UCH\,{\small II}/compact H\,{\small II} regions \citep{Kurtz2005,Martin2005,delaFuente2020,Yang2021}. 
It is to be noted that the presence of extended, diffuse emission could possibly result in large extracted core sizes, leading to underestimation of $n_{\rm e}$ and $EM$. 
Except for the core R5 (due to poor signal-to-noise ratio), the H40$\alpha$ spectra of the other cores are fitted with a single Gaussian profile (see Figure \ref{fig:RGB_vla_ha}(b)). The linewidths are estimated to be in the range $\sim$ [19, 29]$\,\rm km\,s^{-1}$, typical of UCH\,{\small II}/compact H\,{\small II}  \citep{Hoare2007,Yang2021,Liu2021ATOMSIII}.

The inference of the VLA cores representing the intermediate phase between an evolved UCH\,{\small II} region and an early compact H\,{\small II} region is further supported by the identification of radio counterparts from the \textit{CORNISH} survey\footnote{Co-Ordinated Radio 'N' Infrared Survey for High-mass star formation \citep[\textit{CORNISH};][]{Purcell2013}.}. 
Based on the derived radio properties and association with NIR and MIR emission, the classification of the sample of CORNISH UCH\,{\small II} regions is robust and reliable \citep{Kalcheva2018}. 
VLA cores, R1, R4, and R5 are classified as UCH\,{\small II} regions, G024.4721+00.4877, G024.4736+00.4950, and G024.4698+00.4954 \citep{Kalcheva2018}. Core R3 is also detected in the CORNISH survey, but not classified as it does not meet the criteria of flux density greater than $7 \sigma$.

From the ATOMS 3~mm continuum map, we identify ten cores. 
The estimated radii lie in the range $\sim$ [3.7, 9.6]$\times 10^{-2}$ pc with a median value of $7.4\times 10^{-2}$ pc.
The cores, MM1, MM5, and MM10 are co-spatial with radio cores R1, R4, and R6, respectively. Radio core R4 is likely fragmented to MM5 and MM6 in the higher resolution ATOMS continuum map. 
The ATOMS cores, MM3, MM4, and MM9, that do not have radio counterparts, display single-component Gaussian H40$\alpha$ line profiles. The fitted linewidths are in the range $\sim$ [19, 32]$\, \rm km\,s^{-1}$. Given the absence of {\it cm} emission, these could be conjectured as very early stages of massive stars in the HC/UC H{\,\small II} region phase \citep{Liu2021ATOMSIII}. The H40$\alpha$ spectra for cores MM2, MM7, and MM8 have poor signal-to-noise and hence, it is difficult to probe their nature.

With the detection of the ring of bright radio and 3~mm continuum emission harbouring HC, UC, and compact H{\,\small II} regions we are likely witnessing a burst of second epoch of massive star formation. The location of these regions broadly aligns with the theoretical predictions of \citet{Maclow2007}. Here, the authors simulate the dynamical expansion of an H{\,\small II} region into turbulent, self-gravitating gas driven by the ionized gas's overpressure, sweeping up a shell of gas. Theoretically, this shell expands in $\rm \sim 10^5\,yr$ to a radius of $\sim$ 1~pc, and subsequently, the ionized gas breaks out of the natal cloud. The results of this simulation show that an episode of secondary collapse ensues in the shell, where existing turbulent density fluctuations in the shell lead to collapse of self-gravitating cores. Consistent with these predictions, \citet{Hunter2008} presents a case study of the UCH{\,\small II} region G5.89-0.39, where they identified multiple $875\,\mu$m cores confined to the expanding shell formed in the process of secondary collapse.
 
The simulations of \citet{Maclow2007}, however, showed the formation of externally ionized low-mass, transient cores in the shell, masquerading as UCH{\,\small II} regions. In our case, 75\% of the VLA and ATOMS cores are inferred to be in the early phases of HC, UC, or compact H{\,\small II} regions.
From the estimated Lyman continuum photon flux, the compact H{\,\small II} regions are likely ionized by ZAMS stars with spectral type B0$-$O8.5 \citep{Panagia1973} (see Table \ref{tab:param-VLAcores}). Even though the $H-K$ uncertainty allows for the source E4 to be considered as a Class III source, in all likelihood it is a Class II YSO (Figure \ref{fig:cc-cm}(b)) and located within $\sim$3\arcsec, is possibly the ionizing source of the compact H{\,\small II} region, R1.

\subsection{Expanding molecular ring}
A molecular ring is clearly visible from the moment zero maps of H$^{13}$CO$^+$ (1-0) (see Figure 1(d)), HCO+ (see Figure 3(a) - (c)) and the column density map (see Figure \ref{fig:cdmap_molcore}), The expansion of this ring is evident from the investigation of the gas kinematics in Section \ref{sec:vel-struc-ring}.

Twelve molecular cores are identified in this molecular gas ring. The radius, mass, surface mass density, and the virial parameter of the cores lie in the range [0.05, 0.1] pc, [4.3, 30.1] $\rm M_\odot$, [0.1, 0.4] $\rm g\,cm^{-2}$, and [0.7, 3.8], respectively. In the same order, the median values are estimated to be 0.06~pc, 8.8 $\rm M_\odot$, 0.2 $\rm g\,cm^{-2}$, and 1.8. For all the detected molecular cores, the estimated surface mass densities satisfy the threshold of 0.05$\,\rm g\,cm^{-2}$ proposed by \citet{Urquhart2014} for massive star formation. It is worth noting here that in the recent QUARKS (Querying Underlying mechanisms of massive star formation with ALMA-Resolved gas Kinematics and Structures) survey \citep{Liu2024}, nearly half of the identified cores have dense 1.4~mm cold dust counterparts, thus confirming their tendency to form stars.
Gauging their gravitational stability, we find that 50\% of the cores have $\alpha_{\rm vir} < 2$, indicating that these are supercritical and under gravitational collapse in the absence of other supporting mechanisms, such as magnetic fields \citep{Kauffmann2013,Tang2019}. 
The other 50\% are subcritical cores with $\alpha_{\rm vir} > 2$. These are gravitationally unbound and represent transient objects unless one considers other mechanisms, such as magnetic field or external pressure, that would confine these structures \citep{Kauffmann2013,Li2020}. In the case of G24.47, feedback from the newly formed massive stars in the ionized ring and the evidence of the expansion suggest that external pressure will play a key role in confining these compact cores.
We searched for infall signature using the H$^{13}$CO$^+$ (1-0) and HCO$^{+}$ spectra extracted toward these cores, but did not find any conclusive evidence. This could be attributed to the possibility that infall signatures are blended with complex dynamics in the molecular ring, including expanding motions and feedback from stellar winds. Further, in the interferometric observations, we may be missing the total power, which makes it difficult to identify absorption features. Supporting the active star formation activity, we have identified one 70$\mu$m point source within $\sim$ 2.5\arcsec of M12 from the Herschel PACS point source catalog \citep{PACS2020}. 
This finding serves as compelling evidence for the presence of an embedded protostar, as external heating cannot raise temperatures high enough to emit at this wavelength \citep{Stutz2013}.
%
\begin{figure*}
    \includegraphics[width=\linewidth]{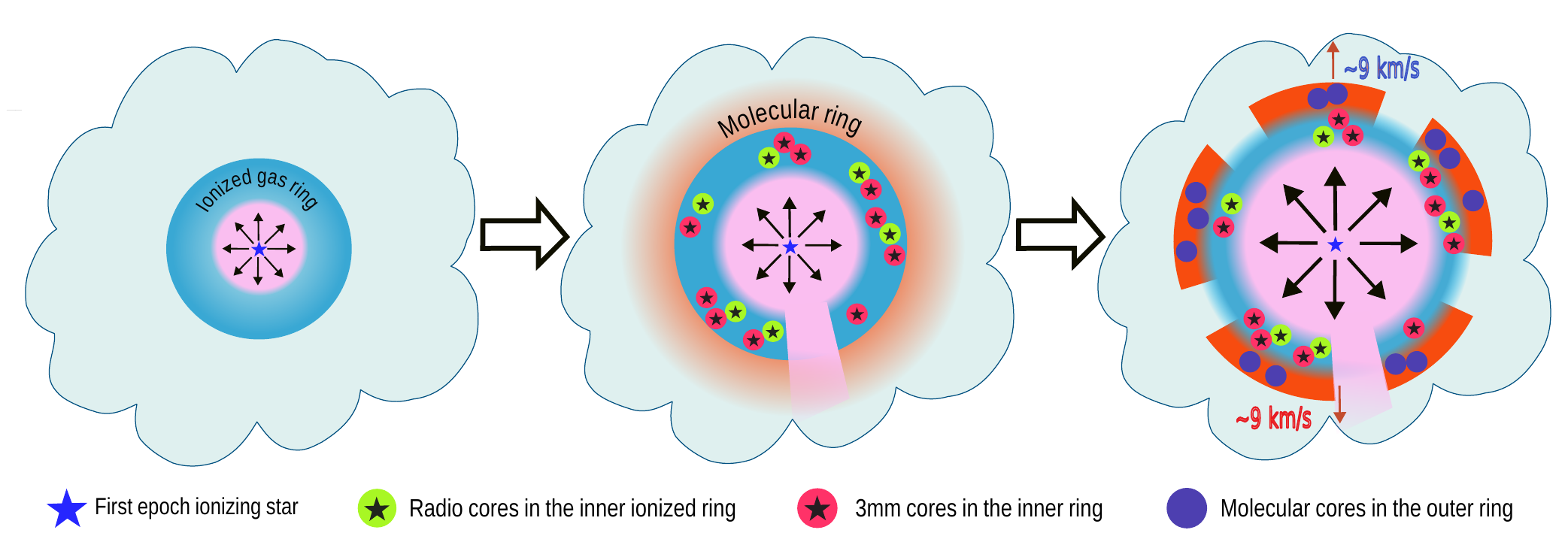}
    \caption{A schematic of the hierarchical triggering of multi-epoch massive star formation in G24.47. }
    \label{fig:schematic}
\end{figure*}
%
\section{Hierarchical triggering of multi-epoch massive star formation}
\label{sec:Hierarchical-trigger-multi-epoch}
Based on our detailed analysis, we propose an interesting picture of multi-epoch massive star formation observed in the G24.47 complex. Our hypothesis is illustrated in the schematic presented in Figure \ref{fig:schematic}. 
The observed hierarchy is initiated with the birth of a massive star in a self-gravitating natal cloud forming an H {\small II} region. The thermal overpressure of the warm ionized gas then drives the expansion of the H {\small II} region. 
This leads to a swept-up ring where secondary collapse occurs, thus triggering the second epoch of star formation. 
In G24.47, we detect a very active high-mass star formation episode in this inner ionized ring with a host of newly formed stars in the early to intermediate evolutionary phases of HC, UC and compact H {\small II} regions with detected RRL and cm emission. 

To support the above inference, we estimate the dynamical timescale of the expanding  H {\small II} region, and compare the same with the typical ages of the HC, UC, and compact {\small II} regions, following the discussion outlined in \citet{Hongliliu2016} and \citet{Swagat2017}.
For an H {\small II} region expanding into a homogeneous medium, the dynamical age is given by \citet{DysonWilliams1980}
\begin{equation}
t_{\rm{dyn}}=\frac{4}{7} \frac{R_{\rm{St}}}{C_{\rm{H} \rm{ II}}}\left[\left(\frac{R_{\rm{IF}}}{R_{\rm{St}}}\right)^{7 / 4}-1\right],
\end{equation}
where
\begin{equation}
R_{\rm{St}}=\left(3 N_{\rm{Ly}} / 4 \pi n_{\rm 0}^2 \alpha_{\rm{B}}\right)^{1 / 3} \\
\end{equation}
In the above equations, ${C_{\rm{H}{II}}}$ is the isothermal sound speed (assumed to be 10$\rm\,km\,s^{-1}$), $R_{\rm{IF}}$ is the radius of the ionized ring ($\sim$0.8 pc), $\alpha_{\rm{B}}$ is the radiative recombination coefficient taken as 2.6$\times 10^{-13} (10^{4} \rm K/T)^{0.7}\, \rm cm^{3} s^{-1}$  \citep{kwan1997}, and $n_{\rm 0}$ is the initial particle density of the ambient gas. We estimate $n_{\rm 0}\sim$ $10^4\,\rm cm^{-3}$ from the retrieved ATLASGAL map (see Appendix \ref{apndx:number-density}), with the assumption that the physical properties, like density, do not change over evolutionary stages (from quiescent to H{\small II} regions) in the star formation process \citep{Urquhart2022}.  Thus, the dynamical age of the H {\small II} region is calculated to be $\rm \sim 2 \times 10^5\,yr$. 
This ensures sufficient time for the formation of the identified UC and compact H {\small II} regions in the inner ionized ring, given the typical lifetimes of these to be $\rm \sim 10^4 - 10^5\,yr$ \citep{Churchwell2002,Davies2011}. 

In recent studies, evidence of expanding ionized ([C {\small II}]) shells has been found in wind-blown bubbles (e.g. Orion Veil; \citealt{Pabst2019}, RCW 120; \citealt{Luisi2021}). Indication of such an expansion of the inner ionized ring in G24.47 is seen from the velocity structure probed with the H40$\alpha$ RRL emission. However, the picture that emerges from the investigation of the molecular gas kinematics reveals, perhaps for the first time, a direct and unambiguous signature of an expanding molecular ring in G24.47.

The total mass of the molecular ring ($M_{\rm shell}$) is estimated to be $ \sim515\rm\,M_\odot$ from the column density map. Taking the expansion velocity ($v_{\rm exp} $) of $\sim9\rm\,km\,s^{-1}$, the kinetic energy ($0.5 M_{\rm shell} v^{2}_{\rm exp}$) of the expanding ring is calculated to be $\sim 4\times 10^{47} \rm erg$. Addressing the energy feedback from the H\,{\small II} region, we estimate the kinetic and thermal energies of the ionized gas to be $\sim6 \times 10^{46} \rm erg$ and $\sim3 \times 10^{46} \rm erg$, respectively, using the expressions from  \citet{Xu2018} and \citet{Li2022}. Individually, these are an order of magnitude lower than the kinetic energy of the molecular shell. 
Our results are similar to those seen in RCW 120 \citep{Luisi2021} and Orion Veil \citep{Pabst2019}, where the authors attribute this to leakage of hot plasma into the surrounding. Indeed, the radio ring in G24.47 is observed to be broken towards the south-west (see Figure \ref{fig:RGB_vla_ha}), which was also reported by \citet{Garay1993}. 

Theoretical studies \citep[e.g.,][]{Haid2018} predict that the energy injection from ionized gas into the surrounding medium would dominate that of the stellar wind. 
To confront the energetics with the creation of the expanding molecular ring in G24.47, we further probe the efficiency of the wind power. Considering a single ionizing O8.5V - O8V star for G24.47, as inferred from the radio flux density, we determine the wind luminosity ($ 3.2\times 10^{35} \dot{M} v^{2}_{\rm w}$; $\dot{M}$ being the mass loss rate in $\rm M_\odot yr^{-1}$ and $v_{\rm w}$ the wind velocity in $\rm km\,s^{-1}$). Using the parameter values for these spectral types from \citet{Martins2017}, we estimate the wind luminosity to be $(4-5)\times 10^{34} \rm erg\,s^{-1}$. This translates to mechanical energy $\sim 3 \times 10^{47} \rm erg$ injected by wind of the ionizing star over the dynamical age of the H\,{\small II} region. This is comparable to the kinetic energy of the molecular gas ring, thus indicating efficient conversion of the mechanical energy of the wind to kinetic energy of the ring driving its expansion. Note that in the above analysis, we have not considered the inclination angle, if any, of the molecular gas ring and the winds from newly formed stars in the ionized ring.

Summarizing the analysis of the energetics, we infer that the total energy budget from the ionizing radiation (both kinetic and thermal) and the stellar wind is sufficient for the creation and expansion of the molecular ring, with the wind kinetic power being possibly the most efficient player.

Furthermore, SO emission, a tracer of low-velocity shocks from H {\small II} regions \citep{Liu2020ATOMSI}, is seen (not presented here) to be co-spatial with the observed H$^{13}$CO$^+$ (1-0) and HCO$^{+}$ (1-0) molecular ring. Coupled with the feedback from the newly formed massive stars, a third epoch of triggered massive star formation is observed in this expanding molecular ring, where potential high-mass star-forming molecular cores are identified. A scenario of `collect and collapse' (CC) 
is evident, where gravitational instabilities result in the fragmentation of swept-up ring to molecular condensations which further fragment to the cores that are observed. Consistent with the prediction of the CC hypothesis \citep{Deharveng2003}, these condensations, in the form of core clusters (i.e., M1--M3, M4--M5, M6--M7, M8--M9, M12, and M10--M11 apparently corresponding to six core clusters, see Figure \ref{fig:RGB_vla_ha}(d)), are observed to be almost {\it regularly} spaced in the molecular gas ring enveloping the inner ionized ring.

\section{Conclusions}
\label{sec:conclusion}
Evidence for triggered star formation linking several epochs of stars around H\,{\small II} regions is difficult to assemble. It is challenging to associate evolved massive stars with the next epoch of star-forming regions, each of which must show indications of ongoing star formation activity. Based on a detailed continuum and multispectral line study of G24.47 using data from the ATOMS survey and archival radio and infrared data, we provide evidence of hierarchical triggering of three epochs of massive star formation.
The first is the massive star(s) responsible for forming the H\,{\small II} region G24.47. Using the 4.86 GHz VLA map, we propose the spectral type to be O8.5V--O8V. 
The inner ring of enhanced radio and 3~mm emission comprises the next epoch of massive stars, which formed due to secondary collapse of the swept-up material. We detected six radio and ten 3~mm cores showing signatures of various evolutionary phases, ranging from the initial stages of gravitational collapse to intermediate phases between UC and compact H {\small II} regions. The molecular gas kinematics analysis unveils direct evidence of an expanding molecular ring powered by the ionized radiation and the stellar wind kinetic energy. 
Furthermore, the molecular gas ring expanding at $\sim$ 9$\,\rm km\,s^{-1}$ hosts the third epoch of potential massive star-forming regions, where we identified twelve molecular cores. Virial analysis indicates that 50\% of them are supercritical and in gravitational collapse.
This observational evidence strongly advocates for more detailed case studies to address the exact influence of expanding H {\small II} regions in triggering further star formation. Detailed kinematic studies of the ionized and neutral material in a sample of promising candidates, utilizing high-resolution data, are essential to understand the underlying physical processes.

\begin{acknowledgments}
The authors thank the referee for insightful comments/suggestions which helped improve the manuscript. 
This work has been supported by the National Key R\&D Program of China (No.\,2022YFA1603101). H.-L. Liu is supported by National Natural Science Foundation of China (NSFC) through the grant No.\,12103045, by Yunnan Fundamental Research Project (grant No.\,202301AT070118, 202401AS070121), and by Xingdian Talent Support Plan -- Youth Project. 
T.L. acknowledges the support by the National Key R\&D Program of China (No. 2022YFA1603101), National Natural Science Foundation of China (NSFC) through grants No.12073061 and No.12122307, the international partnership program of Chinese Academy of Sciences through grant No.114231KYSB20200009, and the Tianchi Talent Program of Xinjiang Uygur Autonomous Region. 
G.G.  and L.B. acknowledge support by the ANID BASAL project FB210003. 
This work was performed in part at the Jet Propulsion Laboratory, California Institute of Technology, under contract with the National Aeronautics and Space Administration (80NM0018D0004). This work is sponsored (in part) by the Chinese Academy of Sciences (CAS), through a grant to the CAS South America Center for Astronomy (CASSACA) in Santiago, Chile. M.J. acknowledges support from the Research Council of Finland grant No. 348342. 
JOC acknowledges financial support from the South African Department of Science and Innovation's National Research Foundation under the ISARP RADIOMAP Joint Research Scheme (DSI-NRF Grant Number 150551). 
This research made use of astrodendro, a Python package to compute dendrograms of Astronomical data ({\url{http://www.dendrograms.org/}}). This research made use of Astropy, a community-developed core Python package for Astronomy \citep{Astropy2018}.
\end{acknowledgments}


\bibliography{reference}{}

\begin{thebibliography}{}
\expandafter\ifx\csname natexlab\endcsname\relax\def\natexlab#1{#1}\fi
\providecommand{\url}[1]{\href{#1}{#1}}
\providecommand{\dodoi}[1]{doi:~\href{http://doi.org/#1}{\nolinkurl{#1}}}
\providecommand{\doeprint}[1]{\href{http://ascl.net/#1}{\nolinkurl{http://ascl.net/#1}}}
\providecommand{\doarXiv}[1]{\href{https://arxiv.org/abs/#1}{\nolinkurl{https://arxiv.org/abs/#1}}}

\bibitem[{{Arce} {et~al.}(2011){Arce}, {Borkin}, {Goodman}, {Pineda}, \& {Beaumont}}]{Arce2011}
{Arce}, H.~G., {Borkin}, M.~A., {Goodman}, A.~A., {Pineda}, J.~E., \& {Beaumont}, C.~N. 2011, \apj, 742, 105, \dodoi{10.1088/0004-637X/742/2/105}

\bibitem[{{Areal} {et~al.}(2020){Areal}, {Buccino}, {Paron}, {Fari{\~n}a}, \& {Ortega}}]{Areal2020}
{Areal}, M.~B., {Buccino}, A., {Paron}, S., {Fari{\~n}a}, C., \& {Ortega}, M.~E. 2020, \mnras, 496, 870, \dodoi{10.1093/mnras/staa1543}

\bibitem[{{Astropy Collaboration} {et~al.}(2018){Astropy Collaboration}, {Price-Whelan}, {Sip{\H{o}}cz}, {G{\"u}nther}, {Lim}, {Crawford}, {Conseil}, {Shupe}, {Craig}, {Dencheva}, {Ginsburg}, {VanderPlas}, {Bradley}, {P{\'e}rez-Su{\'a}rez}, {de Val-Borro}, {Aldcroft}, {Cruz}, {Robitaille}, {Tollerud}, {Ardelean}, {Babej}, {Bach}, {Bachetti}, {Bakanov}, {Bamford}, {Barentsen}, {Barmby}, {Baumbach}, {Berry}, {Biscani}, {Boquien}, {Bostroem}, {Bouma}, {Brammer}, {Bray}, {Breytenbach}, {Buddelmeijer}, {Burke}, {Calderone}, {Cano Rodr{\'\i}guez}, {Cara}, {Cardoso}, {Cheedella}, {Copin}, {Corrales}, {Crichton}, {D'Avella}, {Deil}, {Depagne}, {Dietrich}, {Donath}, {Droettboom}, {Earl}, {Erben}, {Fabbro}, {Ferreira}, {Finethy}, {Fox}, {Garrison}, {Gibbons}, {Goldstein}, {Gommers}, {Greco}, {Greenfield}, {Groener}, {Grollier}, {Hagen}, {Hirst}, {Homeier}, {Horton}, {Hosseinzadeh}, {Hu}, {Hunkeler}, {Ivezi{\'c}}, {Jain}, {Jenness}, {Kanarek}, {Kendrew}, {Kern}, {Kerzendorf}, {Khvalko}, {King}, {Kirkby}, {Kulkarni},
  {Kumar}, {Lee}, {Lenz}, {Littlefair}, {Ma}, {Macleod}, {Mastropietro}, {McCully}, {Montagnac}, {Morris}, {Mueller}, {Mumford}, {Muna}, {Murphy}, {Nelson}, {Nguyen}, {Ninan}, {N{\"o}the}, {Ogaz}, {Oh}, {Parejko}, {Parley}, {Pascual}, {Patil}, {Patil}, {Plunkett}, {Prochaska}, {Rastogi}, {Reddy Janga}, {Sabater}, {Sakurikar}, {Seifert}, {Sherbert}, {Sherwood-Taylor}, {Shih}, {Sick}, {Silbiger}, {Singanamalla}, {Singer}, {Sladen}, {Sooley}, {Sornarajah}, {Streicher}, {Teuben}, {Thomas}, {Tremblay}, {Turner}, {Terr{\'o}n}, {van Kerkwijk}, {de la Vega}, {Watkins}, {Weaver}, {Whitmore}, {Woillez}, {Zabalza}, \& {Astropy Contributors}}]{Astropy2018}
{Astropy Collaboration}, {Price-Whelan}, A.~M., {Sip{\H{o}}cz}, B.~M., {et~al.} 2018, \aj, 156, 123, \dodoi{10.3847/1538-3881/aabc4f}

\bibitem[{{Beaumont} \& {Williams}(2010)}]{Beaumont2010}
{Beaumont}, C.~N., \& {Williams}, J.~P. 2010, \apj, 709, 791, \dodoi{10.1088/0004-637X/709/2/791}

\bibitem[{{Becker} {et~al.}(1994){Becker}, {White}, {Helfand}, \& {Zoonematkermani}}]{Becker1994}
{Becker}, R.~H., {White}, R.~L., {Helfand}, D.~J., \& {Zoonematkermani}, S. 1994, \apjs, 91, 347, \dodoi{10.1086/191941}

\bibitem[{{Benjamin} {et~al.}(2003){Benjamin}, {Churchwell}, {Babler}, {Bania}, {Clemens}, {Cohen}, {Dickey}, {Indebetouw}, {Jackson}, {Kobulnicky}, {Lazarian}, {Marston}, {Mathis}, {Meade}, {Seager}, {Stolovy}, {Watson}, {Whitney}, {Wolff}, \& {Wolfire}}]{Benjamin2003}
{Benjamin}, R.~A., {Churchwell}, E., {Babler}, B.~L., {et~al.} 2003, \pasp, 115, 953, \dodoi{10.1086/376696}

\bibitem[{{Bertoldi} \& {McKee}(1992)}]{Bertoldi1992}
{Bertoldi}, F., \& {McKee}, C.~F. 1992, \apj, 395, 140, \dodoi{10.1086/171638}

\bibitem[{{Bessell} \& {Brett}(1988)}]{Bessel1988}
{Bessell}, M.~S., \& {Brett}, J.~M. 1988, \pasp, 100, 1134, \dodoi{10.1086/132281}

\bibitem[{{Bhadari} {et~al.}(2021){Bhadari}, {Dewangan}, {Zemlyanukha}, {Ojha}, {Zinchenko}, \& {Sharma}}]{Bhadari2021}
{Bhadari}, N.~K., {Dewangan}, L.~K., {Zemlyanukha}, P.~M., {et~al.} 2021, \apj, 922, 207, \dodoi{10.3847/1538-4357/ac2a44}

\bibitem[{{Bonne} {et~al.}(2023){Bonne}, {Kabanovic}, {Schneider}, {Zavagno}, {Keilmann}, {Simon}, {Buchbender}, {G{\"u}sten}, {Jacob}, {Jacobs}, {Kavak}, {Polles}, {Tiwari}, {Wyrowski}, \& {Tielens}}]{Bonne2023}
{Bonne}, L., {Kabanovic}, S., {Schneider}, N., {et~al.} 2023, \aap, 679, L5, \dodoi{10.1051/0004-6361/202347721}

\bibitem[{{Cambr{\'e}sy} {et~al.}(2013){Cambr{\'e}sy}, {Marton}, {Feher}, {T{\'o}th}, \& {Schneider}}]{Cambresy2013}
{Cambr{\'e}sy}, L., {Marton}, G., {Feher}, O., {T{\'o}th}, L.~V., \& {Schneider}, N. 2013, \aap, 557, A29, \dodoi{10.1051/0004-6361/201321235}

\bibitem[{{Churchwell}(2002)}]{Churchwell2002}
{Churchwell}, E. 2002, \araa, 40, 27, \dodoi{10.1146/annurev.astro.40.060401.093845}

\bibitem[{{Churchwell} {et~al.}(1990){Churchwell}, {Walmsley}, \& {Cesaroni}}]{Churchwell1990}
{Churchwell}, E., {Walmsley}, C.~M., \& {Cesaroni}, R. 1990, \aaps, 83, 119

\bibitem[{{Churchwell} {et~al.}(2006){Churchwell}, {Povich}, {Allen}, {Taylor}, {Meade}, {Babler}, {Indebetouw}, {Watson}, {Whitney}, {Wolfire}, {Bania}, {Benjamin}, {Clemens}, {Cohen}, {Cyganowski}, {Jackson}, {Kobulnicky}, {Mathis}, {Mercer}, {Stolovy}, {Uzpen}, {Watson}, \& {Wolff}}]{Churchwell2006}
{Churchwell}, E., {Povich}, M.~S., {Allen}, D., {et~al.} 2006, \apj, 649, 759, \dodoi{10.1086/507015}

\bibitem[{{Contreras} {et~al.}(2016){Contreras}, {Garay}, {Rathborne}, \& {Sanhueza}}]{Contreras2016}
{Contreras}, Y., {Garay}, G., {Rathborne}, J.~M., \& {Sanhueza}, P. 2016, \mnras, 456, 2041, \dodoi{10.1093/mnras/stv2796}

\bibitem[{{Dale} {et~al.}(2015){Dale}, {Haworth}, \& {Bressert}}]{Dale2015}
{Dale}, J.~E., {Haworth}, T.~J., \& {Bressert}, E. 2015, \mnras, 450, 1199, \dodoi{10.1093/mnras/stv396}

\bibitem[{{Das} {et~al.}(2017){Das}, {Tej}, {Vig}, {Liu}, {Liu}, {Ishwara Chandra}, \& {Ghosh}}]{Swagat2017}
{Das}, S.~R., {Tej}, A., {Vig}, S., {et~al.} 2017, \mnras, 472, 4750, \dodoi{10.1093/mnras/stx2290}

\bibitem[{{Davies} {et~al.}(2011){Davies}, {Hoare}, {Lumsden}, {Hosokawa}, {Oudmaijer}, {Urquhart}, {Mottram}, \& {Stead}}]{Davies2011}
{Davies}, B., {Hoare}, M.~G., {Lumsden}, S.~L., {et~al.} 2011, \mnras, 416, 972, \dodoi{10.1111/j.1365-2966.2011.19095.x}

\bibitem[{{de la Fuente} {et~al.}(2020){de la Fuente}, {Tafoya}, {Trinidad}, {Porras}, {Nigoche-Netro}, {Kemp}, {Kurtz}, {Franco}, \& {Rodr{\'\i}guez-Rico}}]{delaFuente2020}
{de la Fuente}, E., {Tafoya}, D., {Trinidad}, M.~A., {et~al.} 2020, \mnras, 497, 4436, \dodoi{10.1093/mnras/staa2149}

\bibitem[{{Deharveng} {et~al.}(2003){Deharveng}, {Lefloch}, {Zavagno}, {Caplan}, {Whitworth}, {Nadeau}, \& {Mart{\'\i}n}}]{Deharveng2003}
{Deharveng}, L., {Lefloch}, B., {Zavagno}, A., {et~al.} 2003, \aap, 408, L25, \dodoi{10.1051/0004-6361:20031157}

\bibitem[{{Deharveng} {et~al.}(2010){Deharveng}, {Schuller}, {Anderson}, {Zavagno}, {Wyrowski}, {Menten}, {Bronfman}, {Testi}, {Walmsley}, \& {Wienen}}]{Deharveng2010}
{Deharveng}, L., {Schuller}, F., {Anderson}, L.~D., {et~al.} 2010, \aap, 523, A6, \dodoi{10.1051/0004-6361/201014422}

\bibitem[{{Dyson} \& {Williams}(1980)}]{DysonWilliams1980}
{Dyson}, J.~E., \& {Williams}, D.~A. 1980, {Physics of the interstellar medium}

\bibitem[{{Elmegreen} \& {Lada}(1977)}]{ElmegreenLada1977}
{Elmegreen}, B.~G., \& {Lada}, C.~J. 1977, \apj, 214, 725, \dodoi{10.1086/155302}

\bibitem[{{Figueira} {et~al.}(2017){Figueira}, {Zavagno}, {Deharveng}, {Russeil}, {Anderson}, {Men'shchikov}, {Schneider}, {Hill}, {Motte}, {M{\`e}ge}, {LeLeu}, {Roussel}, {Bernard}, {Traficante}, {Paradis}, {Tig{\'e}}, {Andr{\'e}}, {Bontemps}, \& {Abergel}}]{Figueira2017}
{Figueira}, M., {Zavagno}, A., {Deharveng}, L., {et~al.} 2017, \aap, 600, A93, \dodoi{10.1051/0004-6361/201629379}

\bibitem[{{Garay} {et~al.}(1993){Garay}, {Rodriguez}, {Moran}, \& {Churchwell}}]{Garay1993}
{Garay}, G., {Rodriguez}, L.~F., {Moran}, J.~M., \& {Churchwell}, E. 1993, \apj, 418, 368, \dodoi{10.1086/173396}

\bibitem[{{Gonz{\'a}lez-Samaniego} \& {Vazquez-Semadeni}(2020)}]{Semadeni2020}
{Gonz{\'a}lez-Samaniego}, A., \& {Vazquez-Semadeni}, E. 2020, \mnras, 499, 668, \dodoi{10.1093/mnras/staa2921}

\bibitem[{{Haid} {et~al.}(2018){Haid}, {Walch}, {Seifried}, {W{\"u}nsch}, {Dinnbier}, \& {Naab}}]{Haid2018}
{Haid}, S., {Walch}, S., {Seifried}, D., {et~al.} 2018, \mnras, 478, 4799, \dodoi{10.1093/mnras/sty1315}

\bibitem[{{Herschel Point Source Catalogue Working Group} {et~al.}(2020){Herschel Point Source Catalogue Working Group}, {Marton}, {Calzoletti}, {Perez Garcia}, {Kiss}, {Paladini}, {Altieri}, {Sanchez Portal}, \& {Kidger}}]{PACS2020}
{Herschel Point Source Catalogue Working Group}, {Marton}, G., {Calzoletti}, L., {et~al.} 2020, {VizieR Online Data Catalog: Herschel/PACS Point Source Catalogs (Herschel team, 2017)}, VizieR On-line Data Catalog: VIII/106. Originally published in: Herschel catalogs (2017)

\bibitem[{{Hoare} {et~al.}(2007){Hoare}, {Kurtz}, {Lizano}, {Keto}, \& {Hofner}}]{Hoare2007}
{Hoare}, M.~G., {Kurtz}, S.~E., {Lizano}, S., {Keto}, E., \& {Hofner}, P. 2007, in Protostars and Planets V, ed. B.~{Reipurth}, D.~{Jewitt}, \& K.~{Keil}, 181, \dodoi{10.48550/arXiv.astro-ph/0603560}

\bibitem[{{Hoq} {et~al.}(2013){Hoq}, {Jackson}, {Foster}, {Sanhueza}, {Guzm{\'a}n}, {Whitaker}, {Claysmith}, {Rathborne}, {Vasyunina}, \& {Vasyunin}}]{Hoq2013}
{Hoq}, S., {Jackson}, J.~M., {Foster}, J.~B., {et~al.} 2013, \apj, 777, 157, \dodoi{10.1088/0004-637X/777/2/157}

\bibitem[{{Hunter} {et~al.}(2008){Hunter}, {Brogan}, {Indebetouw}, \& {Cyganowski}}]{Hunter2008}
{Hunter}, T.~R., {Brogan}, C.~L., {Indebetouw}, R., \& {Cyganowski}, C.~J. 2008, \apj, 680, 1271, \dodoi{10.1086/588016}

\bibitem[{{Kabanovic} {et~al.}(2022){Kabanovic}, {Schneider}, {Ossenkopf-Okada}, {Falasca}, {G{\"u}sten}, {Stutzki}, {Simon}, {Buchbender}, {Anderson}, {Bonne}, {Guevara}, {Higgins}, {Koribalski}, {Luisi}, {Mertens}, {Okada}, {R{\"o}llig}, {Seifried}, {Tiwari}, {Wyrowski}, {Zavagno}, \& {Tielens}}]{Kabanovic2022}
{Kabanovic}, S., {Schneider}, N., {Ossenkopf-Okada}, V., {et~al.} 2022, \aap, 659, A36, \dodoi{10.1051/0004-6361/202142575}

\bibitem[{{Kalcheva} {et~al.}(2018){Kalcheva}, {Hoare}, {Urquhart}, {Kurtz}, {Lumsden}, {Purcell}, \& {Zijlstra}}]{Kalcheva2018}
{Kalcheva}, I.~E., {Hoare}, M.~G., {Urquhart}, J.~S., {et~al.} 2018, \aap, 615, A103, \dodoi{10.1051/0004-6361/201832734}

\bibitem[{{Kauffmann} {et~al.}(2013){Kauffmann}, {Pillai}, \& {Goldsmith}}]{Kauffmann2013}
{Kauffmann}, J., {Pillai}, T., \& {Goldsmith}, P.~F. 2013, \apj, 779, 185, \dodoi{10.1088/0004-637X/779/2/185}

\bibitem[{{Kendrew} {et~al.}(2012){Kendrew}, {Simpson}, {Bressert}, {Povich}, {Sherman}, {Lintott}, {Robitaille}, {Schawinski}, \& {Wolf-Chase}}]{Kendrew2012}
{Kendrew}, S., {Simpson}, R., {Bressert}, E., {et~al.} 2012, \apj, 755, 71, \dodoi{10.1088/0004-637X/755/1/71}

\bibitem[{{Keto} {et~al.}(2008){Keto}, {Zhang}, \& {Kurtz}}]{Keto2008}
{Keto}, E., {Zhang}, Q., \& {Kurtz}, S. 2008, \apj, 672, 423, \dodoi{10.1086/522570}

\bibitem[{{Koornneef}(1983)}]{Koornneef1983}
{Koornneef}, J. 1983, \aap, 128, 84

\bibitem[{{Kurtz}(2005)}]{Kurtz2005}
{Kurtz}, S. 2005, in Massive Star Birth: A Crossroads of Astrophysics, ed. R.~{Cesaroni}, M.~{Felli}, E.~{Churchwell}, \& M.~{Walmsley}, Vol. 227, 111--119, \dodoi{10.1017/S1743921305004424}

\bibitem[{{Kwan}(1997)}]{kwan1997}
{Kwan}, J. 1997, \apj, 489, 284, \dodoi{10.1086/304773}

\bibitem[{{Lada} \& {Adams}(1992)}]{LadaAdams1992}
{Lada}, C.~J., \& {Adams}, F.~C. 1992, \apj, 393, 278, \dodoi{10.1086/171505}

\bibitem[{{Lawrence} {et~al.}(2007){Lawrence}, {Warren}, {Almaini}, {Edge}, {Hambly}, {Jameson}, {Lucas}, {Casali}, {Adamson}, {Dye}, {Emerson}, {Foucaud}, {Hewett}, {Hirst}, {Hodgkin}, {Irwin}, {Lodieu}, {McMahon}, {Simpson}, {Smail}, {Mortlock}, \& {Folger}}]{Lawrence2007}
{Lawrence}, A., {Warren}, S.~J., {Almaini}, O., {et~al.} 2007, \mnras, 379, 1599, \dodoi{10.1111/j.1365-2966.2007.12040.x}

\bibitem[{{Li} {et~al.}(2022){Li}, {Wang}, {Ma}, {Zhang}, {Li}, \& {Zheng}}]{Li2022}
{Li}, C.-X., {Wang}, H.-C., {Ma}, Y.-H., {et~al.} 2022, Research in Astronomy and Astrophysics, 22, 045008, \dodoi{10.1088/1674-4527/ac52a0}

\bibitem[{{Li} {et~al.}(2020){Li}, {Zhang}, {Liu}, {Beuther}, {Palau}, {Girart}, {Smith}, {Hora}, {Lin}, {Qiu}, {Strom}, {Wang}, {Li}, \& {Yue}}]{Li2020}
{Li}, S., {Zhang}, Q., {Liu}, H.~B., {et~al.} 2020, \apj, 896, 110, \dodoi{10.3847/1538-4357/ab84f1}

\bibitem[{{Liu} {et~al.}(2016){Liu}, {Li}, {Wu}, {Yuan}, {Liu}, {Dubner}, {Paron}, {Ortega}, {Molinari}, {Huang}, {Zavagno}, {Samal}, {Huang}, \& {Zhang}}]{Hongliliu2016}
{Liu}, H.-L., {Li}, J.-Z., {Wu}, Y., {et~al.} 2016, \apj, 818, 95, \dodoi{10.3847/0004-637X/818/1/95}

\bibitem[{{Liu} {et~al.}(2021){Liu}, {Liu}, {Evans}, {Wang}, {Garay}, {Qin}, {Li}, {Stutz}, {Goldsmith}, {Liu}, {Tej}, {Zhang}, {Juvela}, {Li}, {Wang}, {Bronfman}, {Ren}, {Wu}, {Kim}, {Lee}, {Tatematsu}, {Cunningham}, {Liu}, {Wu}, {Hirota}, {Lee}, {Li}, {Kang}, {Mardones}, {Ristorcelli}, {Zhang}, {Luo}, {Toth}, {Yi}, {Yun}, {Peng}, {Li}, {Zhu}, {Shen}, {Baug}, {Dewangan}, {Chakali}, {Liu}, {Xu}, {Wang}, {Zhang}, {Li}, {Zhang}, {Zhou}, {Tang}, {Xue}, {Issac}, {Soam}, \& {{\'A}lvarez-Guti{\'e}rrez}}]{Liu2021ATOMSIII}
{Liu}, H.-L., {Liu}, T., {Evans}, Neal~J., I., {et~al.} 2021, \mnras, 505, 2801, \dodoi{10.1093/mnras/stab1352}

\bibitem[{{Liu} {et~al.}(2022){Liu}, {Tej}, {Liu}, {Issac}, {Saha}, {Goldsmith}, {Wang}, {Zhang}, {Qin}, {Wang}, {Li}, {Soam}, {Dewangan}, {Lee}, {Li}, {Liu}, {Zhang}, {Ren}, {Juvela}, {Bronfman}, {Wu}, {Tatematsu}, {Chen}, {Li}, {Stutz}, {Zhang}, {Viktor Toth}, {Luo}, {Xu}, {Li}, {Liu}, {Zhou}, {Zhang}, {Tang}, {Zhang}, {Baug}, {Mannfors}, {Chakali}, \& {Dutta}}]{Liu2022}
{Liu}, H.-L., {Tej}, A., {Liu}, T., {et~al.} 2022, \mnras, 510, 5009, \dodoi{10.1093/mnras/stab2757}

\bibitem[{{Liu} {et~al.}(2023){Liu}, {Tej}, {Liu}, {Sanhueza}, {Qin}, {He}, {Goldsmith}, {Garay}, {Pan}, {Morii}, {Li}, {Stutz}, {Tatematsu}, {Xu}, {Bronfman}, {Saha}, {Issac}, {Baug}, {Toth}, {Dewangan}, {Wang}, {Zhou}, {Lee}, {Yang}, {Luo}, {Shen}, {Zhang}, {Wu}, {Ren}, {Liu}, {Soam}, {Zhang}, \& {Luo}}]{Liu2023}
---. 2023, \mnras, 522, 3719, \dodoi{10.1093/mnras/stad047}

\bibitem[{{Liu} {et~al.}(2020){Liu}, {Evans}, {Kim}, {Goldsmith}, {Liu}, {Zhang}, {Tatematsu}, {Wang}, {Juvela}, {Bronfman}, {Cunningham}, {Garay}, {Hirota}, {Lee}, {Kang}, {Li}, {Li}, {Mardones}, {Qin}, {Ristorcelli}, {Tej}, {Toth}, {Wu}, {Wu}, {Yi}, {Yun}, {Liu}, {Peng}, {Li}, {Li}, {Lee}, {Shen}, {Baug}, {Wang}, {Zhang}, {Issac}, {Zhu}, {Luo}, {Soam}, {Liu}, {Xu}, {Wang}, {Zhang}, {Ren}, \& {Zhang}}]{Liu2020ATOMSI}
{Liu}, T., {Evans}, N.~J., {Kim}, K.-T., {et~al.} 2020, \mnras, 496, 2790, \dodoi{10.1093/mnras/staa1577}

\bibitem[{{Liu} {et~al.}(2024){Liu}, {Liu}, {Zhu}, {Garay}, {Liu}, {Goldsmith}, {Evans}, {Kim}, {Liu}, {Xu}, {Lu}, {Tej}, {Mai}, {Bronfman}, {Li}, {Mardones}, {Stutz}, {Tatematsu}, {Wang}, {Zhang}, {Qin}, {Zhou}, {Luo}, {Zhang}, {Cheng}, {He}, {Gu}, {Li}, {Zhang}, {Zhang}, {Saha}, {Dewangan}, {Sanhueza}, \& {Shen}}]{Liu2024}
{Liu}, X., {Liu}, T., {Zhu}, L., {et~al.} 2024, Research in Astronomy and Astrophysics, 24, 025009, \dodoi{10.1088/1674-4527/ad0d5c}

\bibitem[{{Lockman}(1989)}]{Lockman1989}
{Lockman}, F.~J. 1989, \apjs, 71, 469, \dodoi{10.1086/191383}

\bibitem[{{Lucas} {et~al.}(2008){Lucas}, {Hoare}, {Longmore}, {Schr{\"o}der}, {Davis}, {Adamson}, {Bandyopadhyay}, {de Grijs}, {Smith}, {Gosling}, {Mitchison}, {G{\'a}sp{\'a}r}, {Coe}, {Tamura}, {Parker}, {Irwin}, {Hambly}, {Bryant}, {Collins}, {Cross}, {Evans}, {Gonzalez-Solares}, {Hodgkin}, {Lewis}, {Read}, {Riello}, {Sutorius}, {Lawrence}, {Drew}, {Dye}, \& {Thompson}}]{Lucas2008}
{Lucas}, P.~W., {Hoare}, M.~G., {Longmore}, A., {et~al.} 2008, \mnras, 391, 136, \dodoi{10.1111/j.1365-2966.2008.13924.x}

\bibitem[{{Luisi} {et~al.}(2021){Luisi}, {Anderson}, {Schneider}, {Simon}, {Kabanovic}, {G{\"u}sten}, {Zavagno}, {Broos}, {Buchbender}, {Guevara}, {Jacobs}, {Justen}, {Klein}, {Linville}, {R{\"o}llig}, {Russeil}, {Stutzki}, {Tiwari}, {Townsley}, \& {Tielens}}]{Luisi2021}
{Luisi}, M., {Anderson}, L.~D., {Schneider}, N., {et~al.} 2021, Science Advances, 7, eabe9511, \dodoi{10.1126/sciadv.abe9511}

\bibitem[{{Mac Low} {et~al.}(2007){Mac Low}, {Toraskar}, {Oishi}, \& {Abel}}]{Maclow2007}
{Mac Low}, M.-M., {Toraskar}, J., {Oishi}, J.~S., \& {Abel}, T. 2007, \apj, 668, 980, \dodoi{10.1086/521292}

\bibitem[{{Mart{\'\i}n-Hern{\'a}ndez} {et~al.}(2005){Mart{\'\i}n-Hern{\'a}ndez}, {Vermeij}, \& {van der Hulst}}]{Martin2005}
{Mart{\'\i}n-Hern{\'a}ndez}, N.~L., {Vermeij}, R., \& {van der Hulst}, J.~M. 2005, \aap, 433, 205, \dodoi{10.1051/0004-6361:20042143}

\bibitem[{{Martins} \& {Palacios}(2017)}]{Martins2017}
{Martins}, F., \& {Palacios}, A. 2017, \aap, 598, A56, \dodoi{10.1051/0004-6361/201629538}

\bibitem[{{Meyer} {et~al.}(1997){Meyer}, {Calvet}, \& {Hillenbrand}}]{Meyer1997}
{Meyer}, M.~R., {Calvet}, N., \& {Hillenbrand}, L.~A. 1997, \aj, 114, 288, \dodoi{10.1086/118474}

\bibitem[{{Oey} {et~al.}(2005){Oey}, {Watson}, {Kern}, \& {Walth}}]{Oey2005}
{Oey}, M.~S., {Watson}, A.~M., {Kern}, K., \& {Walth}, G.~L. 2005, \aj, 129, 393, \dodoi{10.1086/426333}

\bibitem[{{Pabst} {et~al.}(2019){Pabst}, {Higgins}, {Goicoechea}, {Teyssier}, {Berne}, {Chambers}, {Wolfire}, {Suri}, {Guesten}, {Stutzki}, {Graf}, {Risacher}, \& {Tielens}}]{Pabst2019}
{Pabst}, C., {Higgins}, R., {Goicoechea}, J.~R., {et~al.} 2019, \nat, 565, 618, \dodoi{10.1038/s41586-018-0844-1}

\bibitem[{{Panagia}(1973)}]{Panagia1973}
{Panagia}, N. 1973, \aj, 78, 929, \dodoi{10.1086/111498}

\bibitem[{{Paron} {et~al.}(2011){Paron}, {Petriella}, \& {Ortega}}]{Paron2011}
{Paron}, S., {Petriella}, A., \& {Ortega}, M.~E. 2011, \aap, 525, A132, \dodoi{10.1051/0004-6361/201015312}

\bibitem[{{Potdar} {et~al.}(2022){Potdar}, {Das}, {Issac}, {Tej}, {Vig}, \& {Chandra}}]{Potdar2022}
{Potdar}, A., {Das}, S.~R., {Issac}, N., {et~al.} 2022, \mnras, 510, 658, \dodoi{10.1093/mnras/stab3479}

\bibitem[{{Purcell} {et~al.}(2009){Purcell}, {Minier}, {Longmore}, {Andr{\'e}}, {Walsh}, {Jones}, {Herpin}, {Hill}, {Cunningham}, \& {Burton}}]{Purcell2009}
{Purcell}, C.~R., {Minier}, V., {Longmore}, S.~N., {et~al.} 2009, \aap, 504, 139, \dodoi{10.1051/0004-6361/200811358}

\bibitem[{{Purcell} {et~al.}(2013){Purcell}, {Hoare}, {Cotton}, {Lumsden}, {Urquhart}, {Chandler}, {Churchwell}, {Diamond}, {Dougherty}, {Fender}, {Fuller}, {Garrington}, {Gledhill}, {Goldsmith}, {Hindson}, {Jackson}, {Kurtz}, {Mart{\'\i}}, {Moore}, {Mundy}, {Muxlow}, {Oudmaijer}, {Pandian}, {Paredes}, {Shepherd}, {Smethurst}, {Spencer}, {Thompson}, {Umana}, \& {Zijlstra}}]{Purcell2013}
{Purcell}, C.~R., {Hoare}, M.~G., {Cotton}, W.~D., {et~al.} 2013, \apjs, 205, 1, \dodoi{10.1088/0067-0049/205/1/1}

\bibitem[{{Quireza} {et~al.}(2006){Quireza}, {Rood}, {Bania}, {Balser}, \& {Maciel}}]{Quireza2006}
{Quireza}, C., {Rood}, R.~T., {Bania}, T.~M., {Balser}, D.~S., \& {Maciel}, W.~J. 2006, \apj, 653, 1226, \dodoi{10.1086/508803}

\bibitem[{{Rosolowsky} {et~al.}(2008){Rosolowsky}, {Pineda}, {Kauffmann}, \& {Goodman}}]{Rosolowsky2008}
{Rosolowsky}, E.~W., {Pineda}, J.~E., {Kauffmann}, J., \& {Goodman}, A.~A. 2008, \apj, 679, 1338, \dodoi{10.1086/587685}

\bibitem[{{Saha} {et~al.}(2022){Saha}, {Tej}, {Liu}, {Liu}, {Issac}, {Lee}, {Garay}, {Goldsmith}, {Juvela}, {Qin}, {Stutz}, {Li}, {Wang}, {Baug}, {Bronfman}, {Xu}, {Zhang}, \& {Eswaraiah}}]{Saha2022}
{Saha}, A., {Tej}, A., {Liu}, H.-L., {et~al.} 2022, \mnras, 516, 1983, \dodoi{10.1093/mnras/stac2353}

\bibitem[{{Sanhueza} {et~al.}(2012){Sanhueza}, {Jackson}, {Foster}, {Garay}, {Silva}, \& {Finn}}]{sanhueza2012}
{Sanhueza}, P., {Jackson}, J.~M., {Foster}, J.~B., {et~al.} 2012, \apj, 756, 60, \dodoi{10.1088/0004-637X/756/1/60}

\bibitem[{{Sanhueza} {et~al.}(2019){Sanhueza}, {Contreras}, {Wu}, {Jackson}, {Guzm{\'a}n}, {Zhang}, {Li}, {Lu}, {Silva}, {Izumi}, {Liu}, {Miura}, {Tatematsu}, {Sakai}, {Beuther}, {Garay}, {Ohashi}, {Saito}, {Nakamura}, {Saigo}, {Veena}, {Nguyen-Luong}, \& {Tafoya}}]{Sanhueza2019}
{Sanhueza}, P., {Contreras}, Y., {Wu}, B., {et~al.} 2019, \apj, 886, 102, \dodoi{10.3847/1538-4357/ab45e9}

\bibitem[{{Saral} {et~al.}(2017){Saral}, {Hora}, {Audard}, {Koenig}, {Mart{\'\i}nez-Galarza}, {Motte}, {Nguyen-Luong}, {Saygac}, \& {Smith}}]{Saral2017}
{Saral}, G., {Hora}, J.~L., {Audard}, M., {et~al.} 2017, \apj, 839, 108, \dodoi{10.3847/1538-4357/aa6575}

\bibitem[{{Schlingman} {et~al.}(2011){Schlingman}, {Shirley}, {Schenk}, {Rosolowsky}, {Bally}, {Battersby}, {Dunham}, {Ellsworth-Bowers}, {Evans}, {Ginsburg}, \& {Stringfellow}}]{Schlingman2011}
{Schlingman}, W.~M., {Shirley}, Y.~L., {Schenk}, D.~E., {et~al.} 2011, \apjs, 195, 14, \dodoi{10.1088/0067-0049/195/2/14}

\bibitem[{{Schmiedeke} {et~al.}(2016){Schmiedeke}, {Schilke}, {M{\"o}ller}, {S{\'a}nchez-Monge}, {Bergin}, {Comito}, {Csengeri}, {Lis}, {Molinari}, {Qin}, \& {Rolffs}}]{Schmiedeke2016}
{Schmiedeke}, A., {Schilke}, P., {M{\"o}ller}, T., {et~al.} 2016, \aap, 588, A143, \dodoi{10.1051/0004-6361/201527311}

\bibitem[{{Skrutskie} {et~al.}(2006){Skrutskie}, {Cutri}, {Stiening}, {Weinberg}, {Schneider}, {Carpenter}, {Beichman}, {Capps}, {Chester}, {Elias}, {Huchra}, {Liebert}, {Lonsdale}, {Monet}, {Price}, {Seitzer}, {Jarrett}, {Kirkpatrick}, {Gizis}, {Howard}, {Evans}, {Fowler}, {Fullmer}, {Hurt}, {Light}, {Kopan}, {Marsh}, {McCallon}, {Tam}, {Van Dyk}, \& {Wheelock}}]{Skrutskie2006}
{Skrutskie}, M.~F., {Cutri}, R.~M., {Stiening}, R., {et~al.} 2006, \aj, 131, 1163, \dodoi{10.1086/498708}

\bibitem[{{Stutz} {et~al.}(2013){Stutz}, {Tobin}, {Stanke}, {Megeath}, {Fischer}, {Robitaille}, {Henning}, {Ali}, {di Francesco}, {Furlan}, {Hartmann}, {Osorio}, {Wilson}, {Allen}, {Krause}, \& {Manoj}}]{Stutz2013}
{Stutz}, A.~M., {Tobin}, J.~J., {Stanke}, T., {et~al.} 2013, \apj, 767, 36, \dodoi{10.1088/0004-637X/767/1/36}

\bibitem[{{Tang} {et~al.}(2019){Tang}, {Koch}, {Peretto}, {Novak}, {Duarte-Cabral}, {Chapman}, {Hsieh}, \& {Yen}}]{Tang2019}
{Tang}, Y.-W., {Koch}, P.~M., {Peretto}, N., {et~al.} 2019, \apj, 878, 10, \dodoi{10.3847/1538-4357/ab1484}

\bibitem[{{Thompson} {et~al.}(2012){Thompson}, {Urquhart}, {Moore}, \& {Morgan}}]{Thompson2012}
{Thompson}, M.~A., {Urquhart}, J.~S., {Moore}, T.~J.~T., \& {Morgan}, L.~K. 2012, \mnras, 421, 408, \dodoi{10.1111/j.1365-2966.2011.20315.x}

\bibitem[{{Urquhart} {et~al.}(2014){Urquhart}, {Moore}, {Csengeri}, {Wyrowski}, {Schuller}, {Hoare}, {Lumsden}, {Mottram}, {Thompson}, {Menten}, {Walmsley}, {Bronfman}, {Pfalzner}, {K{\"o}nig}, \& {Wienen}}]{Urquhart2014}
{Urquhart}, J.~S., {Moore}, T.~J.~T., {Csengeri}, T., {et~al.} 2014, \mnras, 443, 1555, \dodoi{10.1093/mnras/stu1207}

\bibitem[{{Urquhart} {et~al.}(2018){Urquhart}, {K{\"o}nig}, {Giannetti}, {Leurini}, {Moore}, {Eden}, {Pillai}, {Thompson}, {Braiding}, {Burton}, {Csengeri}, {Dempsey}, {Figura}, {Froebrich}, {Menten}, {Schuller}, {Smith}, \& {Wyrowski}}]{Urquhart2018}
{Urquhart}, J.~S., {K{\"o}nig}, C., {Giannetti}, A., {et~al.} 2018, \mnras, 473, 1059, \dodoi{10.1093/mnras/stx2258}

\bibitem[{{Urquhart} {et~al.}(2022){Urquhart}, {Wells}, {Pillai}, {Leurini}, {Giannetti}, {Moore}, {Thompson}, {Figura}, {Colombo}, {Yang}, {K{\"o}nig}, {Wyrowski}, {Menten}, {Rigby}, {Eden}, \& {Ragan}}]{Urquhart2022}
{Urquhart}, J.~S., {Wells}, M.~R.~A., {Pillai}, T., {et~al.} 2022, \mnras, 510, 3389, \dodoi{10.1093/mnras/stab3511}

\bibitem[{{Walch} {et~al.}(2012){Walch}, {Whitworth}, {Bisbas}, {W{\"u}nsch}, \& {Hubber}}]{Walch2012}
{Walch}, S.~K., {Whitworth}, A.~P., {Bisbas}, T., {W{\"u}nsch}, R., \& {Hubber}, D. 2012, \mnras, 427, 625, \dodoi{10.1111/j.1365-2966.2012.21767.x}

\bibitem[{{Walsh} {et~al.}(1998){Walsh}, {Burton}, {Hyland}, \& {Robinson}}]{Walsh1998}
{Walsh}, A.~J., {Burton}, M.~G., {Hyland}, A.~R., \& {Robinson}, G. 1998, \mnras, 301, 640, \dodoi{10.1046/j.1365-8711.1998.02014.x}

\bibitem[{{Wang} {et~al.}(2016){Wang}, {Audard}, {Fontani}, {S{\'a}nchez-Monge}, {Busquet}, {Palau}, {Beuther}, {Tan}, {Estalella}, {Isella}, {Gueth}, \& {Jim{\'e}nez-Serra}}]{Wang2016}
{Wang}, Y., {Audard}, M., {Fontani}, F., {et~al.} 2016, \aap, 587, A69, \dodoi{10.1051/0004-6361/201526637}

\bibitem[{{Watson} {et~al.}(2008){Watson}, {Povich}, {Churchwell}, {Babler}, {Chunev}, {Hoare}, {Indebetouw}, {Meade}, {Robitaille}, \& {Whitney}}]{Watson2008}
{Watson}, C., {Povich}, M.~S., {Churchwell}, E.~B., {et~al.} 2008, \apj, 681, 1341, \dodoi{10.1086/588005}

\bibitem[{{Wink} {et~al.}(1982){Wink}, {Altenhoff}, \& {Mezger}}]{Wink1982}
{Wink}, J.~E., {Altenhoff}, W.~J., \& {Mezger}, P.~G. 1982, \aap, 108, 227

\bibitem[{{Xu} {et~al.}(2023){Xu}, {Wang}, {Liu}, {Goldsmith}, {Zhang}, {Juvela}, {Liu}, {Qin}, {Li}, {Tej}, {Garay}, {Bronfman}, {Li}, {Wu}, {G{\'o}mez}, {V{\'a}zquez-Semadeni}, {Tatematsu}, {Ren}, {Zhang}, {Toth}, {Liu}, {Yue}, {Zhang}, {Baug}, {Issac}, {Stutz}, {Liu}, {Fuller}, {Tang}, {Zhang}, {Dewangan}, {Lee}, {Zhou}, {Xie}, {Jiao}, {Wang}, {Liu}, {Luo}, {Soam}, \& {Eswaraiah}}]{Fengwei2023ATOMSXV}
{Xu}, F.-W., {Wang}, K., {Liu}, T., {et~al.} 2023, \mnras, 520, 3259, \dodoi{10.1093/mnras/stad012}

\bibitem[{{Xu} {et~al.}(2018){Xu}, {Xu}, {Zhang}, {Liu}, {Yu}, {Ning}, \& {Ju}}]{Xu2018}
{Xu}, J.-L., {Xu}, Y., {Zhang}, C.-P., {et~al.} 2018, \aap, 609, A43, \dodoi{10.1051/0004-6361/201629189}

\bibitem[{{Yang} {et~al.}(2021){Yang}, {Urquhart}, {Thompson}, {Menten}, {Wyrowski}, {Brunthaler}, {Tian}, {Rugel}, {Yang}, {Yao}, \& {Mutale}}]{Yang2021}
{Yang}, A.~Y., {Urquhart}, J.~S., {Thompson}, M.~A., {et~al.} 2021, \aap, 645, A110, \dodoi{10.1051/0004-6361/202038608}

\bibitem[{{Zavagno} {et~al.}(2006){Zavagno}, {Deharveng}, {Comer{\'o}n}, {Brand}, {Massi}, {Caplan}, \& {Russeil}}]{Zavagno2006}
{Zavagno}, A., {Deharveng}, L., {Comer{\'o}n}, F., {et~al.} 2006, \aap, 446, 171, \dodoi{10.1051/0004-6361:20053952}

\bibitem[{{Zavagno} {et~al.}(2007){Zavagno}, {Pomar{\`e}s}, {Deharveng}, {Hosokawa}, {Russeil}, \& {Caplan}}]{Zavagno2007}
{Zavagno}, A., {Pomar{\`e}s}, M., {Deharveng}, L., {et~al.} 2007, \aap, 472, 835, \dodoi{10.1051/0004-6361:20077474}

\bibitem[{{Zhang} {et~al.}(2023{\natexlab{a}}){Zhang}, {Zhu}, {Liu}, {Ren}, {Liu}, {Wang}, {Wu}, {Zhang}, {Zhou}, {Tatematsu}, {Garay}, {Tej}, {Li}, {Xu}, {Lee}, {Bronfman}, {Soam}, \& {Li}}]{Zhang2023ATOMSXIV}
{Zhang}, C., {Zhu}, F.-Y., {Liu}, T., {et~al.} 2023{\natexlab{a}}, \mnras, 520, 3245, \dodoi{10.1093/mnras/stad190}

\bibitem[{{Zhang} {et~al.}(2023{\natexlab{b}}){Zhang}, {Wang}, {Liu}, {Zavagno}, {Juvela}, {Liu}, {Tej}, {Stutz}, {Li}, {Bronfman}, {Zhang}, {Goldsmith}, {Lee}, {V{\'a}zquez-Semadeni}, {Tatematsu}, {Jiao}, {Xu}, {Wang}, \& {Zhou}}]{Zhang2023ATOMSXIII}
{Zhang}, S., {Wang}, K., {Liu}, T., {et~al.} 2023{\natexlab{b}}, \mnras, 520, 322, \dodoi{10.1093/mnras/stad011}

\end{thebibliography}
\bibliographystyle{aasjournal}
%
%
\appendix
\restartappendixnumbering
%
\section{Channel map}
{The channel map of HCO$^{+}$ is shown in Figure \ref{fig:hco-chan-map}.
\begin{figure*}
    \centering
    \includegraphics[width=0.85\textwidth]{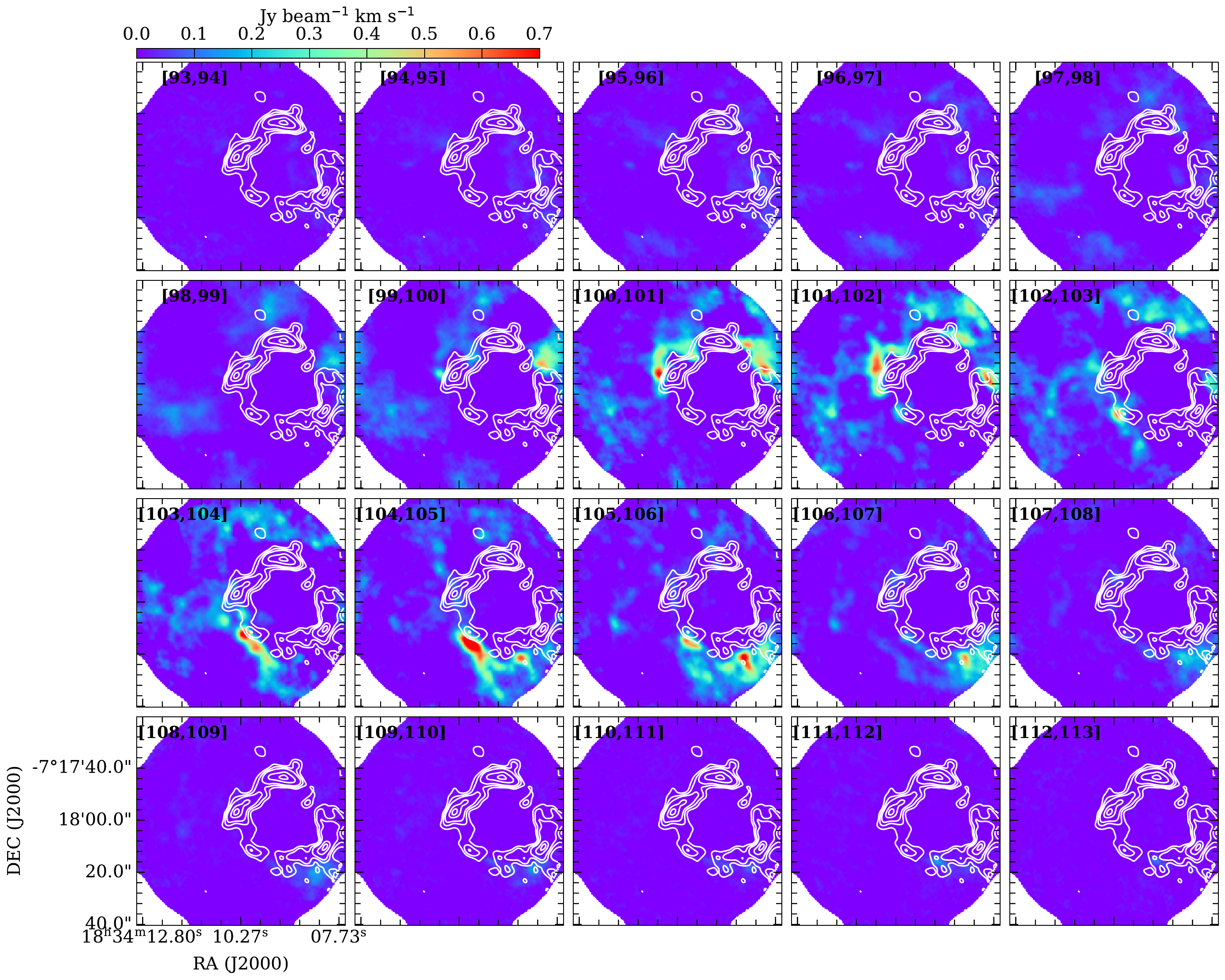}
    \caption{Channel maps of HCO$^{+}$ overlaid with H40$\alpha$ contours (Gaussian smoothed over 5 pixels) with levels starting at 2$\sigma$ ( $\sigma$ = 0.04 $\rm Jy\,beam^{-1}\,km\,s^{-1}$ ) in steps of 1$\sigma$. Each panel shows the velocity integrated intensity within a velocity range of 1 $\rm km\,s^{-1}$. The velocity range is mentioned in the top left of each panel.}
    \label{fig:hco-chan-map}
\end{figure*}
%
%
\section{Core extraction}
\label{apndx:core_extract}
%
\begin{table*}
\caption{Parameters of VLA cores.}
\setlength{\tabcolsep}{2.5pt}
\centering
\begin{tabular}{c c  c c c c c c c c c c c}
 \hline
\multirow{2}{*}{Core} & \multicolumn{2}{c}{Peak position}  & Decon. size & $R_{\rm core}^{\rm VLA}$ & $F^{\rm 4.86}_{\rm int}$ &  $N_{\rm ly}$ & EM & $n_{\rm e}$ & $M_{\rm ion}^{a}$ & $V^{b}_{\rm LSR}$ & $\Delta V^{b}$ & Spec. type\\ [0.5ex] 
& RA(J2000) & DEC(J2000) & ($^{''}\times ^{''}$) & (pc) & ($\rm mJy$) & ($10^{47}\,\rm s^{-1}$)  & ($\rm 10^{6}cm^{-6}pc$) & ($\rm 10^{3}cm^{-3}$) &($\rm M_\odot$) & ($\rm km\,s^{-1}$) & ($\rm km\,s^{-1}$) &\\ [0.5ex] 
 \hline
R1 & 18:34:10.33 & -7:17:55.38 & 13.8 $\times$ 9.6 & 0.16 &383.5  & 14.1 & 2.3 & 2.7 & 1.2 & 106.4 & 24.3 \bla & O9--O8.5 \\
\hline
R2 & 18:34:09.81 & -7:18:11.53 & 10.5 $\times$ 7.6 & 0.13 &200.6 &  7.4 & 2.0 & 2.8 & 0.6 & 101.8 & 21.2 \bla & O9.5--O9 \\
\hline
R3 & 18:34:09.15 & -7:18:12.56 & 12.2 $\times$ 8.8 & 0.15  & 207.1 &  7.6 & 1.6 & 2.3 & 0.7 & 115.8 & 18.9 \bla & O9.5--O9 \\
\hline
R4 & 18:34:09.08 & -7:17:44.06 & 13.6 $\times$ 7.4 & 0.14  & 308.9 &  11.4 & 2.5 & 3.0 & 0.9 & 100.4 & 28.6 \bla & O9.5--O9 \\
\hline
R5 & 18:34:08.51 & -7:18:48.71 & 10.6 $\times$ 6.6 & 0.12  &186.1 &  6.8 & 2.2 & 3.0 & 0.5 & $-^b$ & $-^b$  & B0--O9.5 \\
\hline
R6 & 18:34:08.02 & -7:18:03.72 & 13.8 $\times$ 7.9 & 0.15  &293.8 &  10.8 & 2.2 & 2.7 & 0.9 & 108.8 & 27.6 \bla  & O9.5--O9\\
\hline
\end{tabular}
\label{tab:param-VLAcores}
\begin{flushleft}
{\bf Note:} $^a$Mass of ionised gas, $M_{\rm ion} = \frac{4}{3}\pi \big( R_{\rm core}^{\rm VLA}\big)^3 n_{\rm e} m_{\rm p}$. $^b$ $V_{\rm LSR}$ and $\Delta V$ are obtained from ATOMS H40$\alpha$ transition. For R5, the spectrum is not fitted with Gaussian as it shows a high signal-to-noise ratio (see Figure \ref{fig:RGB_vla_ha}(b)).
\end{flushleft}
\end{table*}
%
\begin{table*}
\caption{Parameters of detected cores from ALMA 3~mm map.}
\centering
\begin{tabular}{c c c c c c  c c c}
 \hline\
\multirow{2}{*}{Core} & \multicolumn{2}{c}{Peak position}  & Decon. size &  $R^{\rm a}_{\rm core}$  & $F^{\rm 3mm}_{\rm peak}$ & $F^{\rm 3mm}_{\rm int}$ & V$_{\rm LSR}$ & $\Delta V^{b}$ \\ [0.5ex] 
& RA(J2000) & DEC(J2000) & ($^{''} \times {}^{''}$) & ($10^{-2}$ pc) & ($\rm mJy\,beam^{-1}$) & ($\rm mJy$) & ($\rm km\,s^{-1}$) & ($\rm km\,s^{-1}$)   \\ [0.5ex] 
 \hline
MM1 &18:34:10.35   & -7:17:55.88 & 8.8 $\times$ 5.2 &  9.6&  8.3 &  111.8 & 105.7 & 24.9\\
\hline
MM2 &18:34:10.14   & -7:18:06.45 & 3.4 $\times$ 2.7 &  4.0&  4.5 &  15.5  & $-^c$ & $-^c$\\
\hline
MM3 & 18:34:10.03  & -7:18:11.31 & 5.9 $\times$ 4.0 &  6.8&  5.2 &  38.2  & 98.1 & 23.6\\
\hline
MM4 & 18:34:09.69  & -7:18:14.13 & 6.1 $\times$ 5.4 &  8.0&  4.0 &  38.6 & 98.1 & 19.4\\
\hline
MM5 &18:34:09.07   & -7:17:44.44 & 5.7 $\times$ 3.0 &  5.8&  7.6 &  43.1  & 101.1 & 25.7\\
\hline
MM6 & 18:34:09.16  & -7:17:42.02 & 2.9 $\times$ 2.3 &  3.7&  7.6 &  20.7  & 99.6 & 27.3\\
\hline
MM7 & 18:34:08.42  & -7:17:51.85 & 7.4 $\times$ 4.5 &  8.0&  5.3 &  53.5  & $-^c$ & $-^c$\\
\hline
MM8 & 18:34:08.31  & -7:18:18.31 & 4.5 $\times$ 3.9 &  5.9&  3.7 &  20.9  & $-^c$ & $-^c$\\
\hline
MM9 & 18:34:08.09  & -7:18:09.27 & 7.2 $\times$ 4.3 &  7.9&  6.9 &  64.8  & 116.0 & 32.2 \\
\hline
MM10 & 18:34:08.08 & -7:18:02.61 & 7.6 $\times$ 4.8 &  8.5&  6.8 &  73.2  & 106.5 & 23.4 \\ 
\hline
\end{tabular}
\label{tab:parameter3mm}
\begin{flushleft}
{\bf Note:}
$^{\rm a}$$R_{\rm core}$ is the core radius taken to be half of the geometric mean of $\rm FWHM_{maj}$ and $\rm FWHM_{min}$ at the core distance. $^b$$\Delta V$ is derived from ATOMS H40$\alpha$ transition. $^c$Spectra of MM2, MM7 and MM8 have poor signal-to-noise ratio.
\end{flushleft}
\end{table*}
\begin{table*}
\caption{Parameters of molecular cores.}
\centering
\begin{tabular}{c c c c c c c c c c c c}
 \hline
\multirow{2}{*}{Core} & \multicolumn{2}{c}{Peak position} & Decon. Size  & $R_{\rm eff}^{\rm mol^a}$ & $\Delta V$ & N$(\rm H_2)^b$ & $n_{\rm H_2}^c$& $M_{\rm core}^{\rm mol}$ & $\Sigma^d$ & $M_{\rm vir}$ & $\alpha_{\rm vir}$\\ [0.5ex] 
& RA(J2000) & DEC(J2000) & (${}^{''} \times {}^{''}$) & ($10^{-2}$ pc) & ($\rm km\,s^{-1}$) & ($10^{22}\rm cm^{-2}$) & ($10^{5}\rm cm^{-2}$) & ($\rm M_\odot$) & ($\rm g\,cm^{-2}$) & ($\rm M_\odot$) & \\ [0.5ex] 
 \hline
M1 & 18:34:10.79    & -07:18:01.24 & 5.8 $\times$ 3.7 & 6.5 & 1.2 & 3.1 & 1.2 & 9.2 & 0.1 & 14.4 & 1.4 \\
\hline
M2 & 18:34:10.79    & -07:17:50.95 & 4.3 $\times$ 2.7 & 4.7 & 1.3 & 5.3 & 2.6 & 8.3 & 0.2 & 12.4	& 1.3 \\
\hline
M3 & 18:34:10.71	& -07:17:55.09 & 5.9 $\times$ 3.3 & 6.2 & 1.2 & 5.8 & 2.2 & 15.7 & 0.3 & 12.5 & 0.7 \\
\hline
M4 & 18:34:10.14	& -07:18:13.52 & 5.0 $\times$ 2.8 & 5.3 & 2.5 & 8.4 & 3.9 & 16.2 & 0.4 & 48.2 & 2.5 \\
\hline
M5 & 18:34:09.90	& -07:18:16.81 & 6.2 $\times$ 3.8 & 6.9 & 1.5 & 8.8 & 3.1 & 28.6 & 0.4 & 22.5 & 0.7 \\
\hline
M6 & 18:34:09.65	& -07:17:30.47 & 4.2 $\times$ 3.3 & 5.3 & 2.0 & 4.1 & 1.8 & 7.7 & 0.2 & 32.4 &3.8 \\
\hline
M7 & 18:34:09.29	& -07:17:28.14 & 9.4 $\times$ 5.6 & 10.2 & 2.5 & 4.2 & 1.0 & 30.1 & 0.2 & 93.8 & 2.7 \\
\hline
M8 & 18:34:08.68	& -07:17:40.75 & 5.6 $\times$ 2.5 & 5.2 & 1.2 & 3.1 &  1.5 & 6.1  & 0.2 & 10.6 & 1.4 \\
\hline
M9 & 18:34:08.50	& -07:17:43.08 & 4.9 $\times$ 3.5 & 5.8 & 1.3 & 3.5 & 1.5 & 8.4  & 0.2 & 14.2 & 1.5 \\
\hline
M10 &18:34:08.56	& -07:18:21.89 & 5.5 $\times$ 3.3 & 6.0 & 1.6 & 4.3 & 1.0 & 6.3 & 0.1 & 22.2 & 3.0 \\
\hline
M11 &18:34:08.36	& -07:18:25.68 & 5.2 $\times$ 2.2 & 4.8 & 1.3 & 2.6 & 1.3 & 4.3 & 0.1 & 11.7 & 2.1 \\
\hline
M12 &18:34:07.99	& -07:17:53.00 & 4.8 $\times$ 4.0 & 6.2 & 2.1 & 4.6 & 1.8 & 12.5 & 0.2 & 38.5 & 2.9 \\
\hline
\end{tabular}
\label{tab:molcore-param}
\begin{flushleft}
{\bf Note:} $^a$$R_{\rm eff}^{\rm mol}$ is the effective radius of core equals half of the geometric mean of major and minor axes at the source distance. $^b$Mean value of N$(\rm H_2)$. $^c$Number density ($n_{\rm H_2} = 3M_{\rm core}^{\rm mol} /4 \pi (R_{\rm eff}^{\rm mol})^3 \mu_{H_2} m_{H})$. $^d$ Mass surface density ($\Sigma = M_{\rm core}/\pi (R_{\rm eff}^{\rm mol^a})^2 $).
\end{flushleft}
\end{table*}
\bla 
Following the approach used in \citet{Li2020,Liu2021ATOMSIII,Liu2022,Saha2022}, cores are extracted by utilizing the \texttt{ASTRODENDRO} package and the \texttt{CASA} imfit task. 
Initially, in the dendrogram algorithm \citep{Rosolowsky2008}, for extraction of the radio and 3~mm cores, we set $min\_value = 3\sigma$ and $min\_delta = \sigma$, where $\sigma$ represents the {\it rms} noise of the corresponding map.
The $min\_npix$ parameter is set to be equivalent to the synthesized beam area for extracting the 3~mm cores and half of the synthesized beam area to detect the radio cores. 
Subsequently, the \texttt{CASA} {\it imfit} task is employed with the \texttt{DENDROGRAM} parameter estimates as initial guess values. In our study, we only concentrated on the structures located on the ring. To ensure the exclusion of spurious cores, we retain only the cores with peak flux greater than 5$\sigma$. Further, we also reject cores with poorly fitted shapes by visual inspection of the map overlaid with the identified structures and discarded structures located at the edge that appeared truncated.
This process identified six radio (R1 - R6) and ten 3~mm cores (MM1 - MM10) which are shown as ellipses in Figure \ref{fig:RGB_vla_ha}(b) and (c), respectively. 
It is to be noted that in the case of 3~mm cores MM5 and MM6, which are also identified in \citet{Liu2021ATOMSIII}, they are manually fitted using \texttt{CASA} {\it imfit} since the \texttt{DENDROGRAM} algorithm resolved it into a single leaf though two distinct cores are seen visually.

In the case of the molecular cores, we employed the same procedure on the moment zero map of H$^{13}$CO$^+$ (see Figure \ref{fig:RGB_vla_ha}(d)) taken over the velocity range of 93.0 to 113.0 $\rm km\,s^{-1}$. Ten cores (M2 - M9 and M11 - M12) are identified. Two additional cores (M1 and M10) are extracted from the column density map obtained using H$^{13}$CO$^+$ and HCO$^+$ (see Appendix \ref{apndx:column_density}). We only concentrated on the structures on the molecular gas ring encompassing the ionized ring. 
The spatial distribution of the detected molecular cores overlaid on the moment zero map of H$^{13}$CO$^+$ and the column density map are illustrated in Figures \ref{fig:RGB_vla_ha}(d) and \ref{fig:cdmap_molcore}, respectively.
%
%
\begin{figure}
    \centering
    \includegraphics[width=\linewidth]{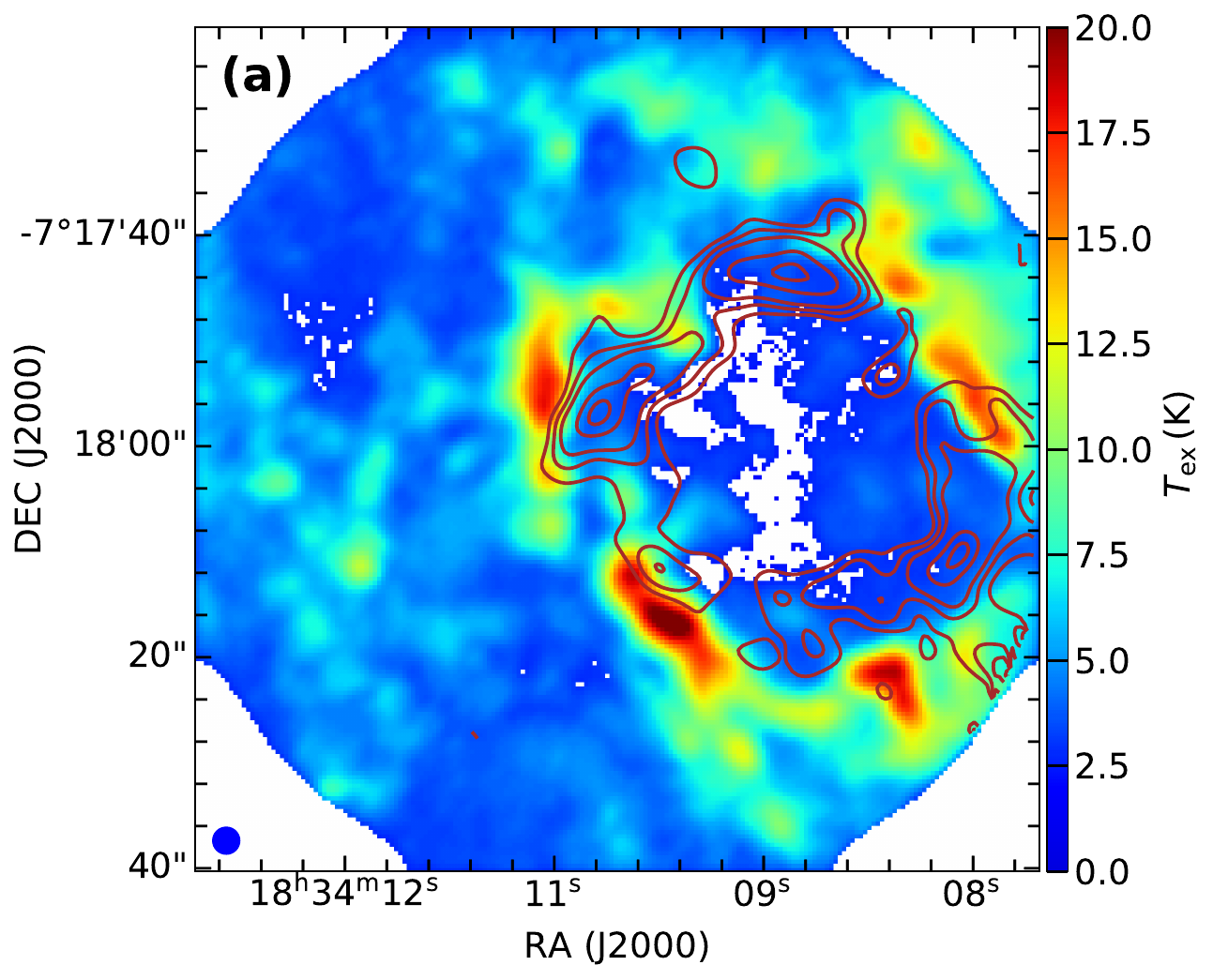}
    \includegraphics[width=0.995\linewidth]{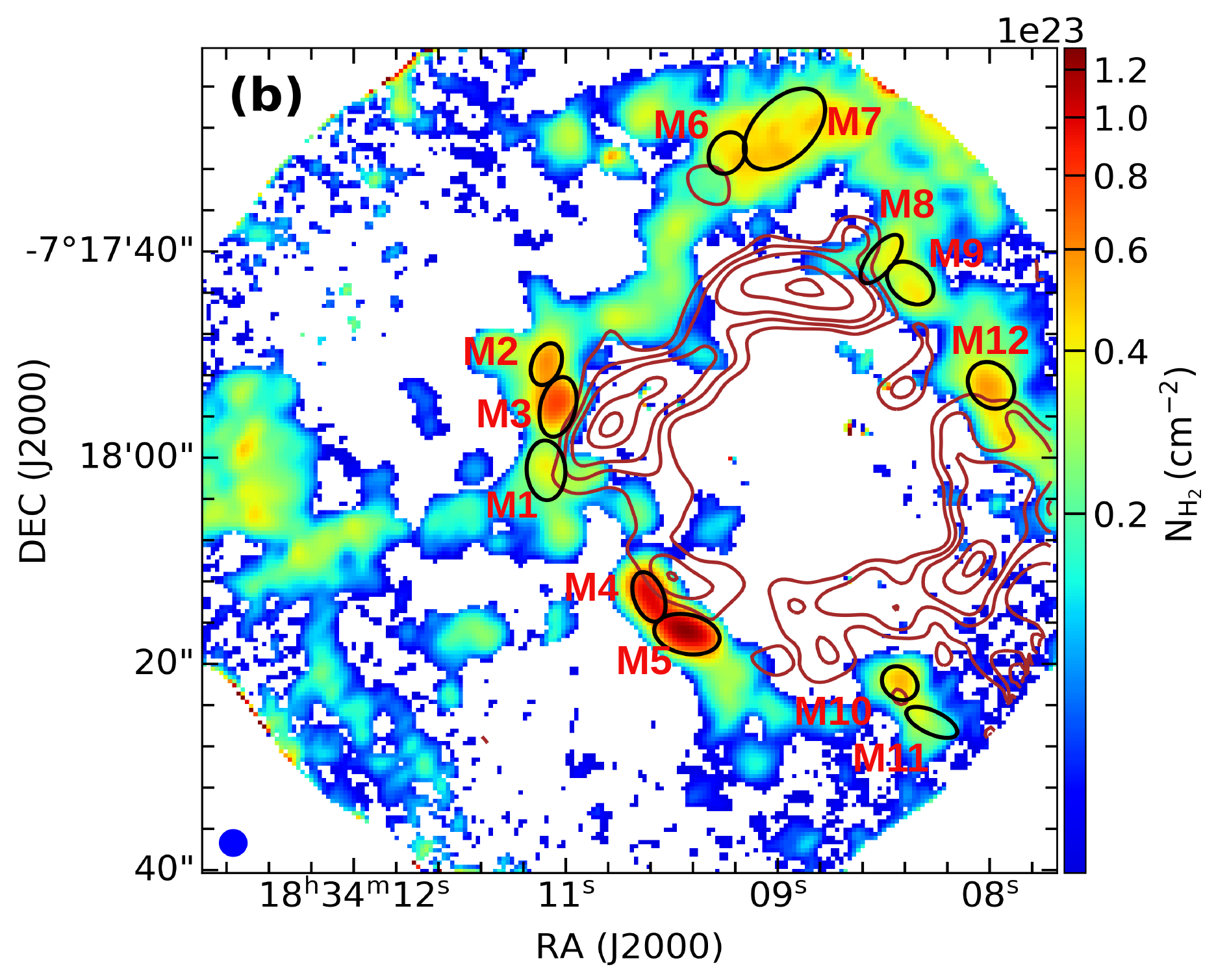}
    \caption{(a) Effective line of sight excitation temperature map generated using HCO$^+$. (b) Column density map towards G24.47, generated using molecular transitions H$^{13}$CO$^+$ and HCO$^+$. The ellipses represent the apertures of identified molecular cores. In both panels, the overlaid contours show the H40$\alpha$ emission with contour levels starting at 2$\sigma$ ( $\sigma =$ 0.04 $\rm Jy\,beam^{-1}\,km\,s^{-1}$) in steps of 1$\sigma$. These contours are smoothed over 5 pixels using Gaussian kernel.  The beam size of 2.5\arcsec is indicated at the bottom left. }
    \label{fig:cdmap_molcore}
\end{figure}
\section{Column density}
\label{apndx:column_density}
Following the method discussed in \citet{Fengwei2023ATOMSXV}, we generate the column density map. Firstly, we assumed HCO$^+$ to be optically thick (optical depth, $\tau \gg 1$), and determined the effective line of sight excitation temperature ($T_{\rm ex}$) for each pixel in the HCO$^+$ spectral cube, following the discussion given in Appendix F of \citet{Fengwei2023ATOMSXV}.
For our analysis, we considered only the pixels where the peak intensity along the spectral axis is greater than four times \textit{rms}. We convolved the data cubes of HCO$^+$ and H$^{13}$CO$^+$ to a common beam size of 2.5\arcsec and also assumed that the two molecules share the same $T_{\rm ex}$. 
Further considering the excitation temperature to be equal to the kinetic temperature for all the energy states and the levels populated according to Boltzmann distribution, the H$^{13}$CO$^+$ column density at each pixel is calculated using Equation 11 of \citet{Fengwei2023ATOMSXV}. 
Further, we have adopted an abundance ratio of H$^{13}$CO$^+$ to H$_2$ of $1.28 \times 10^{-10}$, as determined by \citet{Hoq2013} in their investigation of 333 high-mass star-forming regions using MALT90 data. The generated excitation temperature and hydrogen column density maps are presented in Figure \ref{fig:cdmap_molcore} (a) and (b), respectively. 
\begin{table}
\caption{Candidate ionizing sources.}
\centering
\setlength{\tabcolsep}{4pt}
\begin{tabular}{c c c c c c}
 \hline
\multirow{2}{*}{Source} & \multicolumn{2}{c}{Coordinates} & $J $ & $H$ & $K$ \\ [0.5ex] 
& RA(J2000) & DEC(J2000) & (mag) & (mag) & (mag) \\ [0.5ex] 
 \hline
E1$^\dag$ & 18:34:10.55 & -7:18:14.80 & 15.748 & 14.563 & 13.878 \\ 
\hline
E2$^\star$ & 18:34:10.49 & -7:18:00.37 & 13.021 & 11.853 & 11.135 \\
\hline
E3$^\star$ & 18:34:10.37 & -7:17:53.11 & 13.771 & 12.313 & 11.489 \\
\hline
E4$^\star$ & 18:34:10.26 & -7:17:58.62 & 14.555 & 12.966 & 11.852 \\
\hline
E5$^\dag$ & 18:34:09.98 & -7:18:16.60 & 14.265 & 13.291 & 12.683 \\
\hline
E6$^\dag$ & 18:34:09.91 & -7:18:04.12 & 14.602 & 13.586 & 12.833 \\
\hline
E7$^\dag$ & 18:34:09.52 & -7:18:06.55 & 13.984 & 13.030 & 12.401 \\
\hline
E8$^\star$ & 18:34:09.32 & -7:18:09.18 & 11.547 & 10.356 & 9.603 \\
\hline
E9$^\dag$ & 18:34:09.31 & -7:17:45.66 & 14.634 & 13.263 & 12.456 \\
\hline
E10$^\dag$ & 18:34:09.22 & -7:17:56.34 & 15.592 & 14.503 & 13.879 \\
\hline
E11$^\dag$ & 18:34:09.02 & -7:17:59.67 & 16.188 & 14.521 & 13.525 \\
\hline
E12$^\dag$ & 18:34:08.75 & -7:18:06.21 & 16.338 & 14.260 & 13.101 \\
\hline
\end{tabular}
\label{tab:ionisingstar-param}
\begin{flushleft}
{\bf Note:} $^\star$ and $^\dag$ correspond to values from the 2MASS and UKIDSS catalog, respectively. 
\end{flushleft}
\end{table}
\section{Ionizing massive star(s)}
\label{apndx:Ionizing_star}
To search for candidate ionizing star(s), we select a region covering the observed radio emission. The NIR colour-magnitude and colour-colour plots for the selected region are shown in Figures \ref{fig:cc-cm} (a) and (b).
UKIDSS data has saturation limits of 12.65, 12.5, and 12 mag in $J$, $H$, and $K$, respectively \citep{Lucas2008}. Hence, for sources brighter than this, 2MASS data are used. To ensure that the retrieved sample are sources with good quality photometry, we include 2MASS sources with ``read-flag''=2, and UKIDSS sources with ``pstar'' $>$ 0.94 and ``cl'' = -1. To account for the zero-point photometric offset between the two data sets used, we adopt the approach used by \citet[e.g,][]{Saral2017}.
For this, we consider a large (radius of 2.5\arcmin) region centered on G24.47 and estimate the median and standard deviation of the photometric offset between UKIDSS and 2MASS for sources detected in both data sets. Median values of 0.09, -0.06, and -0.03 are calculated for $J$, $H$, and $K$ bands, respectively, with a standard deviation of 0.1 in each band. Subsequently, we apply an offset of 0.09$\pm$0.1, -0.06$\pm$0.1, -0.03$\pm$0.1 mag, respectively, to the $J$, $H$, and $K$ magnitudes of UKIDSS. 

From the colour-magnitude diagram, we identify 23 massive stars earlier than the B3 spectral type. Next, we compare their location with that of Class III sources.
Taking the photometric and offset uncertainties into account, of the 23 identified stars, 12 lie within the reddening vectors of B3 and O5. These are labelled E1 - E12 and the locations of these are shown in Figure \ref{fig:cc-cm} (c) and the details of these sources are listed in Table \ref{tab:ionisingstar-param}.
\begin{figure*}
    \centering
    \includegraphics[width=\linewidth]{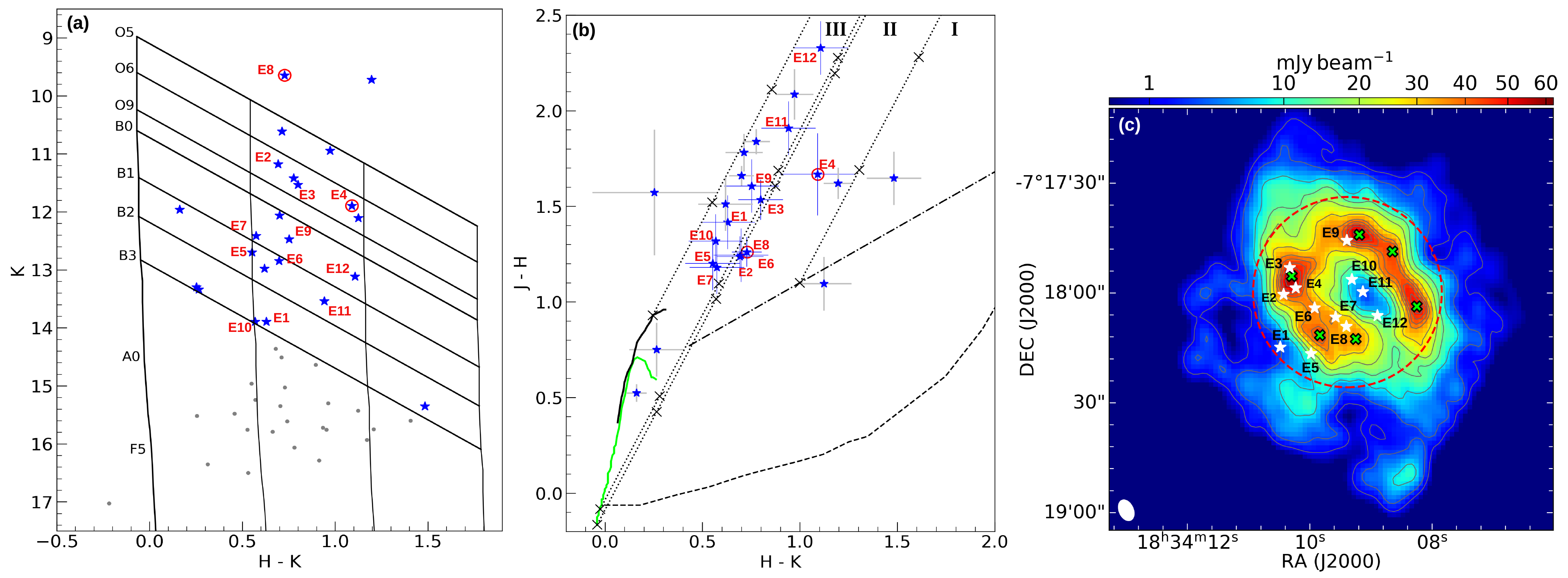}
    \caption{(a) The colour-magnitude diagram of the sources associated with G24.47 within the circle shown in (c). 
    The ZAMS loci, with the spectral types indicated, are drawn for 0, 10, 20, and 30 magnitudes of visual extinction corrected for the distance. The parallel lines are the reddening vectors for the spectral types. Stars earlier than spectral type B3 are shown as stars.  
    (b) The colour-colour diagram for the 23 sources earlier than B3. The loci of giants and main sequence stars, taken from \citet{Koornneef1983} and \citet{Bessel1988}, are shown as black and green curves, respectively. The dashed line shows the locus of the Herbig AeBe stars adopted from \citet{LadaAdams1992}. The locus of classical T Tauri \citep{Meyer1997} is shown as dash-dot line. The parallel lines are the reddening vectors where the ones corresponding to M4, B3 and O5, are shown as dotted lines. The cross marks indicate intervals of 5 mag of visual extinction. 
    Sources (E1 - E12) within the reddening vectors of B3 and O5 are labeled in panels (a) and (b). Both plots are in the \citet{Bessel1988} system. 
    (c) The colour scale shows the VLA 4.86 GHz map of the region associated with G24.47 with contour levels at 3, 7, 20, 35, 50, 70, 90, 100, 110, 120, 135 times $\sigma$ ( $\sigma =$ 0.4 $\rm mJy\,beam^{-1}$). The crosses (‘X’) represent the central positions of the VLA cores. The positions of the sources E1 - E12 are indicated. The ellipse at the bottom left shows the beam. }
    \label{fig:cc-cm}
\end{figure*}
\section{Initial particle number density}
\label{apndx:number-density}
The particle number density, $n_{\rm 0}$, is obtained by estimating the mass of the material within the ionized ring (radius of $\sim$0.8~pc). For this, we used the ATLASGAL 870 $\mu$m map for the region associated with G24.47.  The mass, $M$, is given by
\begin{equation}
M = \frac{F_{\rm \nu}\,d^{2}\,R_{\rm gd}}{B_{\nu}(T_{\rm d})\,\kappa_{\nu}}, 
\end{equation}
where $F_{\nu}$ is the integrated 870$\mu$m flux density; $d$ is the distance to the source; $R_{\rm gd}$ is the gas to dust ratio taken as 100 and $B_{\nu}$ is the Planck function at dust temperature $T_{\rm d}$ taken as 30.4~K \citep{Liu2020ATOMSI}; ${\kappa}_{\nu}$ is the dust opacity coefficient taken to be 1.85 $\rm cm^2\,g^{-1}$ \citep{Urquhart2018}. This gives a mass estimate of 1409 $\rm M_{\odot}$ and particle number density of $n_{\rm 0}\sim$ $10^4\,\rm cm^{-3}$.
%
%
%
%

\end{document}